\documentclass[12pt]{article}
\usepackage{a4wide}
\usepackage{cite}
\usepackage{hyperref}
\usepackage{graphicx}
\graphicspath{{./}{./figures/}}

\newcommand{\be}{\beta}
\newcommand{\sbz}{  }
\newcommand{\prd}{\partial}
\newcommand{\asy}{\mbox{\scriptsize asymp}}
\newcommand{\obe}{\ovl{\beta}}
\newcommand{\beos}{\beta^{\mathrm{OS}}}
\newcommand{\al}{\alpha}
\newcommand{\oal}{\overline{\alpha}}
\newcommand{\oalfp}{\left(\frac{\overline{\alpha}}{4\,\pi}\right)}
\newcommand{\alp}{\left(\frac{\alpha}{\pi}\right)}
\newcommand{\oalpMmu}{\left(\frac{\alpha_{\mu}}{\pi}\right)}
\newcommand{\LMQ}{\ell_{MQ}}
\newcommand{\ellMQ}{\ell_{MQ}}
\newcommand{\lmuQ}{\ell_{\mu Q}}
\newcommand{\ovl}[1]{\overline{#1}}
\newcommand{\re}[1]{(\ref{#1})}
\newcommand{\beq}{\begin{equation}}
\newcommand{\eeq}{\end{equation}}
\newcommand{\bea}{\begin{eqnarray}}
\newcommand{\eea}{\end{eqnarray}}
\newcommand{\dmu}{\mu^2\frac{d}{d\mu^2}}
\newcommand{\MSbar}{\overline{\mbox{MS}}}
\newcommand{\msbar}{{\scriptsize \overline{\rm MS}}}

\newcommand{\BreakI}{ \right. \nonumber \\ &{}& \left. }
\newcommand{\vep}{\varepsilon}
\newcommand{\z}[1]{\zeta_{#1}}
\newcommand{\Pa}[1]{a_{#1}}
\newcommand{\amu}[2]{a_{\mu}^{#1,\{#2\}}}
\newcommand{\Log}[2]{\ln^{\!#2}\!{(#1)}}
\newcommand{\logtwo}{\ln\!{(2)}}
\newcommand{\lmusdmes}{\ell_{{\mu}e}}
\newcommand{\ph}{\phantom{-}}
\let\*=\,
\begin{document}    

\begin{titlepage}
\noindent
\hfill MPP-2012-105\\
\mbox{}
\hfill SFB/CPP-12-40\\
\mbox{}
\hfill TTP12--12-021
\mbox{}

\vspace{0.5cm}
\begin{center}
  \begin{Large}
    \begin{bf}
      The relation between the QED charge renormalized\\[-0.15cm]
      in $\MSbar$ and on-shell schemes at four loops,\\[-0.15cm]
      the QED on-shell {\boldmath{$\beta$}}-function at five loops and\\[-0.15cm]
      asymptotic contributions to the muon anomaly\\
      at five and six loops
    \end{bf}
  \end{Large}
  \vspace{0.8cm}

  \begin{large}
    P.A. Baikov$\rm \, ^{a \,}$, 
    K.G. Chetyrkin$\rm \, ^{b,}$\footnote{Permanent address: Institute
    for Nuclear Research, Russian Academy of Sciences, Moscow 117312, Russia.},
    J.H. K{\"u}hn$\rm \, ^{b \,}$
    {\normalsize and } C. Sturm$\rm \, ^{c \,}$
  \end{large}
  \vskip .7cm
	 {\small {\em 
 	     $\rm ^a$ 
Skobeltsyn Institute of Nuclear Physics, Lomonosov Moscow State
University, \\ 1(2), Leninskie gory, Moscow 119234, Russian Federation
             }}
	 \vskip .3cm
	{\small {\em 
	    $\rm ^b$ 
	    Institut f{\"u}r Theoretische Teilchenphysik,
            Universit{\"a}t Karlsruhe,\\
            D-76128 Karlsruhe, Germany}}
	 \vskip .3cm
	{\small {\em 
	    $\rm ^c$ 
            Max-Planck-Institut f{\"u}r Physik 
            (Werner-Heisenberg-Institut),
            F{\"o}hringer Ring 6,\\
            D-80805 M{\"u}nchen, Germany}}
	
	\vspace{0.8cm}
{\bf Abstract}
\end{center}
\begin{quotation}
  \noindent
  In this paper we compute the four-loop corrections to the QED photon
  self-energy $\Pi(Q^2)$ in the two limits of $q = 0$ and $Q^2 \to
  \infty$. These results are used to explicitly construct the conversion
  relations between the QED charge renormalized in on-shell(OS) and $\MSbar$
  scheme. Using these relations and results of Baikov et al. \cite{Baikov:2012zm} we
  construct the momentum dependent part of $\Pi(Q^2,m,\alpha)$ at
  large $Q^2$ at five loops in both $\MSbar$ and OS schemes.  As a direct
  consequence we arrive at the full result for the QED $\beta$-function
  in the OS scheme at five loops. These results are applied, in turn, to
  analytically evaluate a class of asymptotic contributions to the muon
  anomaly at five and six loops.
\end{quotation}
\end{titlepage}
%
%
%
%
\section{Introduction\label{sec:Introduction}}
\begin{figure}[!ht]
\begin{center}
\begin{minipage}{2cm}
\includegraphics[bb=108 561 223 687,width=2cm]{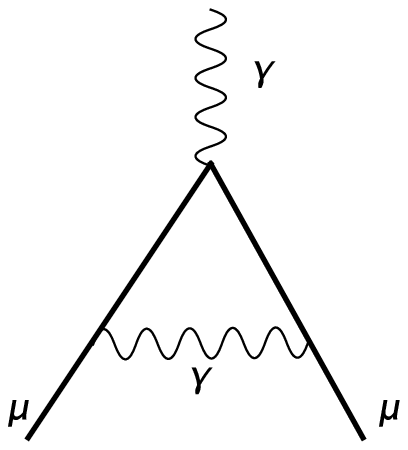}\\[-1cm]
\begin{center}
\vspace{-0.25cm}
{{\tiny$(a)$}\\[-0.25cm]\tiny{pure QED}}
\end{center}
\end{minipage}\hspace{0.5cm}
\begin{minipage}{5cm}
\begin{minipage}{2cm}
\includegraphics[bb=108 561 223 687,width=2cm]{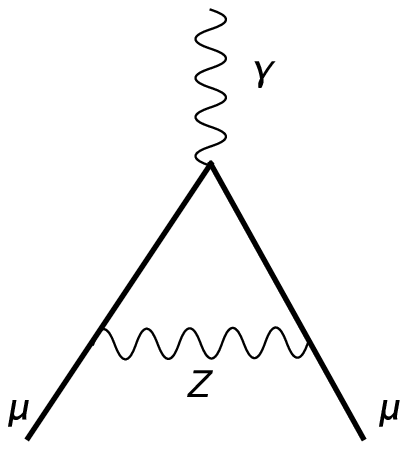}\\[-1.2cm]
\begin{center}
{\tiny$(b_1)$}
\end{center}
\end{minipage}
\hspace{0.5cm}
\begin{minipage}{2cm}
\includegraphics[bb=108 561 223 687,width=2cm]{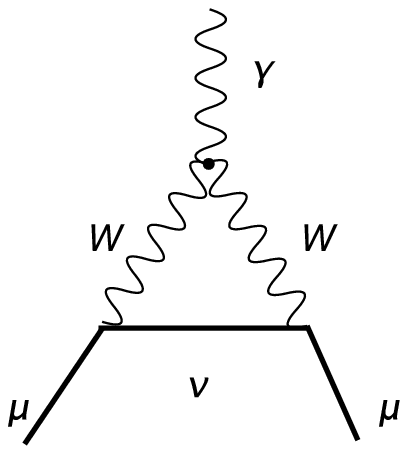}\\[-1.2cm]
\begin{center}
{\tiny$(b_2)$}
\end{center}
\end{minipage}
\vspace{-0.4cm}
\begin{center}
{\tiny{electroweak}}
\end{center}
\end{minipage}\hspace{0.5cm}
\begin{minipage}{2cm}
\includegraphics[bb=108 561 223 687,width=2cm]{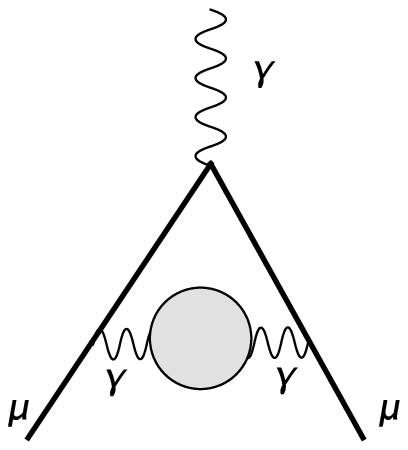}\\[-1.2cm]
\begin{center}
\vspace{-0.1cm}
{\tiny$(c)$}\\[-0.25cm]{\tiny{hadronic}}
\end{center}
\end{minipage}
\end{center}
\vspace{-0.5cm}
\caption{\label{fig:classes} Sample  diagrams for
  the classification into: pure QED contributions {\small{$(a)$}}, electroweak
  corrections {\small{$(b_1)$, $(b_2)$}}, and hadronic contributions
  {\small{$(c)$}}.}
\end{figure}

Quantum electrodynamics (QED) is one of the best tested and established 
quantum field theories. The study of its input parameters and properties
is thus a challenge for each theoretician and experimentalist. For
example the anomalous magnetic moment of the muon $a_\mu$ has been
measured with impressing accuracy at the level of 0.5 parts per
million~\cite{Bennett:2006fi,Nakamura:2010zzi}: $a^{\mbox{\tiny
    exp}}_\mu=116592089(63) \cdot 10^{-11}$. Also from theory side the
anomalous magnetic moment has been studied in great detail through the
computation of higher order corrections. In general higher order
corrections to $a^{theo}_\mu$ are classified into three classes: pure
QED, electroweak and hadronic contributions, where sample  diagrams
are depicted in Fig.~\ref{fig:classes}.

\noindent
Within this work we consider higher order corrections to the pure QED
part. Recently the complete tenth order QED contribution has been
determined numerically in Ref.~\cite{Aoyama:2012wk}. We refer to the reviews
\cite{Melnikov:2006sr,Jegerlehner:2008zza,Jegerlehner:2009ry} and references therein
for a discussion of the lower order QED corrections, the electroweak 
and hadronic contributions.

\begin{figure}[!ht]
\begin{center}
\begin{minipage}{2cm}
\includegraphics[bb=108 562 224 687,width=2cm]{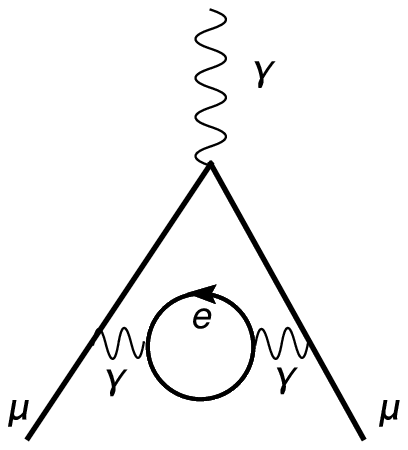}\\[-1.2cm]
\begin{center}
{\tiny{eVP}}
\end{center}
\end{minipage}
\hspace{1cm}
\begin{minipage}{2.2cm}
\includegraphics[bb=88 559 240 687,width=2.2cm]{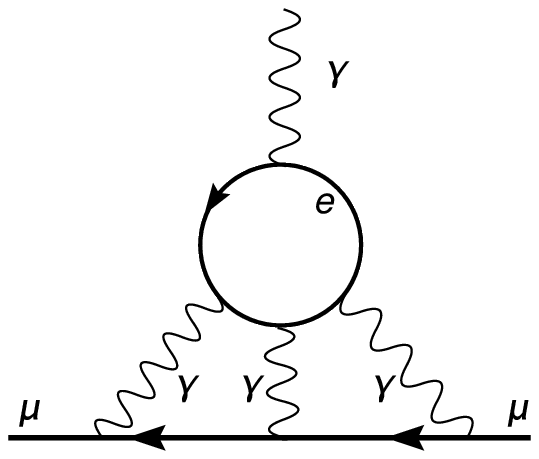}\\[-1.0cm]
\begin{center}
{\tiny{LBL}}
\end{center}
\end{minipage}
\end{center}
\vspace{-0.5cm}
\caption{\label{fig:VPLBL}Example diagrams leading to dominant
  logarithmic contributions from electron vacuum polarization
  insertions and light-by-light scattering diagrams.}
\end{figure}

In QED diagrams with internal fermion loops arise starting from two-loop order,
where the fermion type of the internal loop can be in general different 
from the external muon. In the case that the internal fermion loop
consists of an electron,  logarithmic contributions of the type
$\ln(M_\mu/M_e)$ arise, where $M_\mu$ is the mass of the muon and $M_e$
the mass of the electron. In view of the large mass ratio
$M_\mu/M_e\sim 200$ one expects these logarithms to play a
dominant role. These logarithmically enhanced contributions arise on the one
hand from the insertion of the electron vacuum polarization (eVP)
into the first order muon vertex diagram, shown in Fig.~\ref{fig:classes}(a),
but they can on the other hand also appear through light-by-light (LBL) scattering
diagrams. Examples for both diagram types are shown in Fig.~\ref{fig:VPLBL}.

\noindent
It has been demonstrated in Ref.~\cite{Lautrup:1974ic}, that the
asymptotic part of the anomalous magnetic moment of the muon, $a^{\scriptsize asymp}_\mu$,
which contains these logarithmic contributions originating from the
electron vacuum polarization function insertions, as well as the mass
independent term can be obtained with the help of the electron
vacuum polarization function in the asymptotic limit $M_e\to0$. This technique has
been applied in order to derive $a^{\scriptsize asymp}_\mu$ up to
three-loop 
and four-loop order in
Refs.~\cite{Lautrup:1974ic,Broadhurst:1992za,Kinoshita:1990ur,Baikov:1995ui}.
A similar technique has also been applied in Ref.~\cite{Barbieri:1974nc}.

At five-loop level all diagrams contributing to $a^{theo}_\mu$ can be
decomposed into six supersets, which can be further subdivided into 32
gauge invariant subsets\cite{Nio:2007zz}. These supersets have been computed
numerically in Ref.~\cite{Aoyama:2012wk}. For a few subsets also
analytical results were obtained in
Refs.~\cite{Kataev:1991cp,Laporta:1994md}. 

In this work we compute analytically the asymptotic limit of  the photon propagator with massive electron loops 
and photon exchanges  at four-loop order in two renormalization schemes: 
the $\MSbar$-one   and the  classical on-shell scheme. The calculation is then employed for the 
derivation of  the $\MSbar$--on-shell relation of the fine structure constant
$\alpha$ at four-loop order. This relation allows the conversion of
observables, conveniently computed in the $\MSbar$ scheme, into the on-shell(OS)
scheme or vice versa. These conversion relations  combined with the recently published QED $\beta$-function
in the $\MSbar$ scheme at  five loops \cite{Baikov:2012zm} are then used to  derive the complete 
five-loop contribution to the QED  $\beta$-function in the OS scheme as
well as the {\em momentum-dependent} part of the
polarization function in both the $\MSbar$ and OS scheme at five loops.  
These new four-loop and five-loop results are subsequently used  to determine
analytically $a^{\scriptsize asymp}_\mu$ at five loops as well as some genuinely six-loop
contributions to $a^{\scriptsize asymp}_\mu$. This will constitute 
a new analytical result for several gauge invariant subclasses of the five-loop
QED muon anomaly and will serve as a check for some already known numerical
results. 

The outline of the paper is as follows: in
Section~\ref{sec:GeneralNotations} we introduce our notations and
conventions. In Section~\ref{sec:VacPol} we discuss the methods of
calculation and present the results for the vacuum polarization function
for small and large $Q^2$ in $\MSbar$ and on-shell scheme. In the next
Section we use these results to derive the conversion formulas between
the QED charge, renormalized in $\MSbar$ and on-shell schemes. In
Section~\ref{sec:PolOp5} the polarization operator at five loops is
discussed. In the following Section~\ref{sec:betaOS} we derive the complete
five-loop contribution to the QED $\beta$-function in the OS scheme as
well as the {\em momentum-dependent} part of the polarization function in
both $\MSbar$ and OS schemes at five loops.  In
Sections~\ref{sec:muon5} and \ref{sec:muon6} we analytically compute
{\em all} asymptotic contributions to the muon anomaly at five loops and
logarithmically enhanced ones at six loops respectively.  Finally in
Section~\ref{sec:DiscussConclude} we close with a brief summary and our
conclusions.

\section{Notation and generalities\label{sec:GeneralNotations}}
\begin{figure}[!ht]
\begin{center}
\begin{minipage}{3.5cm}
\includegraphics[bb=106 599 259 700,width=3.5cm]{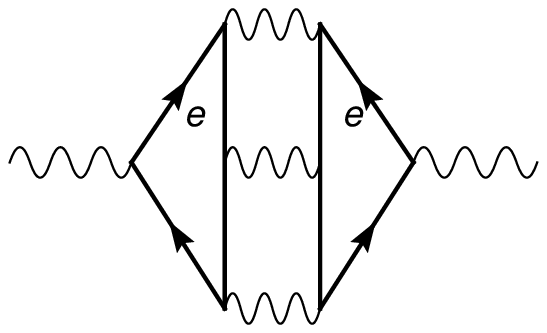}\\[-1cm]
\begin{center}
\vspace{-0.2cm}
{\tiny$(s)$}
\end{center}
\end{minipage}\hspace{0.4cm}
\begin{minipage}{3.5cm}
\includegraphics[bb=106 601 259 697,width=3.5cm]{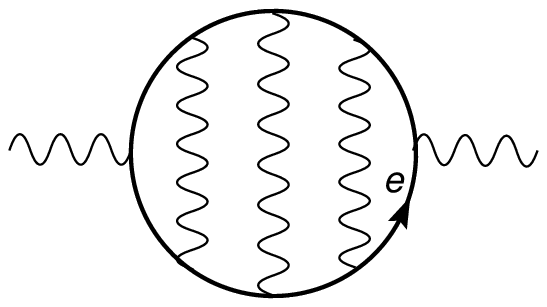}
\begin{center}
\vspace{-0.7cm}
{\tiny$(a)$}
\end{center}
\end{minipage}\hspace{0.4cm}
\begin{minipage}{3.5cm}
\includegraphics[bb=106 601 259 697,width=3.5cm]{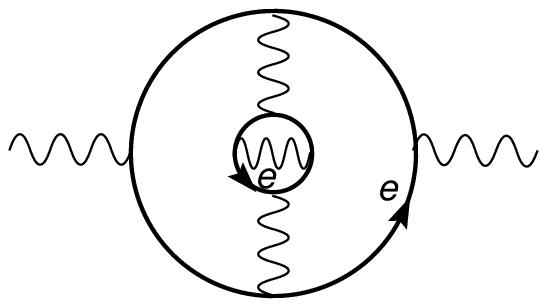}
\begin{center}
\vspace{-0.7cm}
{\tiny$(b)$}
\end{center}
\end{minipage}\hspace{0.4cm}
\begin{minipage}{3.5cm}
\includegraphics[bb=106 601 259 697,width=3.5cm]{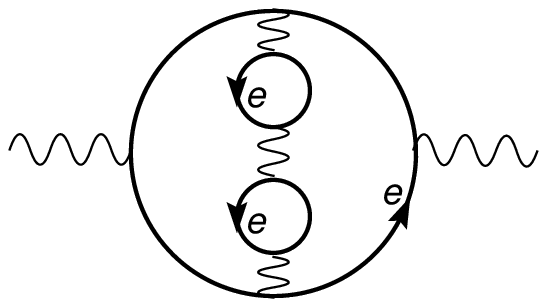}
\begin{center}
\vspace{-0.7cm}
{\tiny$(c)$}
\end{center}
\end{minipage}
\end{center}
\vspace{-0.5cm}
\caption{\label{fig:VPclassification}The four diagram classes arising in
  the computation of the four-loop vacuum polarization function are
  shown. Only one representative for each class is given.  Class $(s)$
  are singlet type diagrams, whereas class $(a),(b)$ and $(c)$ are
  non-singlet type diagrams, which are decomposed according to the number of
  inserted fermion loops.}
\end{figure}

We consider QED with $N$ identical fermions with mass $m$. The $\MSbar$
renormalized photon propagator $\ovl{D}^{\mu\nu}_{R}$ is related to the
photon polarization function $\ovl{\Pi}(-q^{2}/\mu^2,\ovl{m}/\mu,\ovl{\alpha})$ in
the standard way:
\beq
\ovl{D}^{\mu\nu}_{R}(-q^{2},\ovl{m},\ovl{\alpha},\mu) = \frac{-
  g_{\mu\nu}}{-q^2} \frac{1}{1+\ovl{\Pi}(\frac{-q^{2}}{\mu^2},\frac{\ovl{m}}{\mu},\ovl{\alpha})}
 +q^\mu q^\nu \ \  \mbox{terms}
\label{photon_prop}
{}, 
\eeq 
where $\ovl{\alpha} \equiv \alpha^{\msbar}(\mu)$,  $\ovl{m} \equiv
\ovl{m}^{\msbar}(\mu)$ and $\mu$ are the running coupling constant, the fermion mass
and normalization scale in $\MSbar$ scheme respectively. The symbol $q$ denotes the external
Minkowskian momentum of the polarization function.
At four-loop order the appearing self-energy diagrams can be classified
into four classes: singlet and non-singlet type diagrams, where the
non-singlet ones can be again decomposed with respect to the number of
inserted closed fermion loops. Some example diagrams are shown in
Fig.~\ref{fig:VPclassification}.
\\
As is well-known  the $\MSbar$
renormalized photon self-energy does not have any power-not suppressed terms of the 
type $\ln(\ovl{m}^2)$ in the limit of large $Q^2 \equiv -q^2$.
They appear only at order $\mathcal{O}(\ovl{m}^4/Q^4)$ of the large Q asymptotic expansion 
(see, e.g. Ref.~\cite{Chetyrkin:1996ia}). This allows  to define the {\em asymptotic}
part    of the polarization function $\ovl{\Pi}$ in the limit $\ovl{m} \to 0 $ as follows:
\beq
\ovl{\Pi}^{\mbox{\scriptsize asymp}}\bigg(\frac{-q^{2}}{\mu^2}, \ovl{\alpha}\bigg) 
\equiv
\ovl{\Pi}\bigg(\frac{Q^{2}}{\mu^2},\frac{\ovl{m}}{\mu} =0,\ovl{\alpha}\bigg)
\label{asymp:msbar}
{}.
\eeq

The running of the  coupling constant $\ovl{\al}(\mu)$ is governed by the corresponding evolution equation of the form:
\beq
\dmu \ln \ovl{\al} =  \ovl{\beta} (\ovl{\al})
\label{beta:ms:def}
{}.
\eeq
The evolution equation for the   polarization function can  be directly obtained from the fundamental concept 
of the {\em scheme-invariant} charge \cite{Bogolyubov:1956gh,Shirkov:1998ak}, defined through
\begin{equation}
\label{eq:InvCharge}
{\ovl{\alpha} \over 1+   \ovl{\Pi}(\frac{-q^{2}}{\mu^2},\frac{\ovl{m}}{\mu},\ovl{\alpha})}
{}.
\end{equation}
For the case of the asymptotic polarization function the evolution
equation assumes the following, particularly simple form:
\beq
\Biggl\{
\mu^2 \frac{\prd }{\prd \mu^2} + \ovl{\beta}(\oal)\,\left( \oal\, \frac{\prd }{\prd \oal} -1 \right) 
\Biggr\}\,(1+\ovl{\Pi}^{\mbox{\scriptsize asymp}}(Q^{2}/\mu^2, \ovl{\alpha}))  =0
\label{PiMS:RG:massless}
{}.
\eeq
Recently the QED $\beta$-function, as defined in the
    $\MSbar$-scheme, has been evaluated to five-loop order~\cite{Baikov:2012zm}. 
For convenience of the reader we provide the corresponding result below\footnote{Note, please, that the function 
 $\ovl{\beta} (\ovl{\al})$  is related to $\beta^{QED}(A)$ as given in Eq.~(4.7) of \cite{Baikov:2012zm} as  
follows: $\ovl{\beta} (\ovl{\al}) \equiv \left(\beta^{QED}(A)/A\right)|_{A = \ovl{\al}/(4\pi)}$.
}:
\bea
&{}& \ovl{\beta}(\oal) = 
\frac{4\,N}{3}\frac{\oal}{4\,\pi}  
{+}  
4\,  N \oalfp^2
-  \oalfp^3
\left[
2  \,N 
+\frac{44}{9}  \, N^2
\right]
\nonumber\\
&{+}&  \oalfp^4
\left[
-46  \,N 
+\frac{760}{27}  \, N^2
-\frac{832}{9}  \,\zeta_{3} \, N^2
-\frac{1232}{243}  \, N^3
\right]
\nonumber\\
&{+}& \oalfp^5 \Biggl( \,N 
\left[
\frac{4157}{6} 
+128  \sbz \zeta_{3}
\right]
{+} \, N^2
\left[
-\frac{7462}{9} 
-992  \sbz \zeta_{3}
+2720  \sbz \zeta_{5}
\right]
\nonumber\\
&{+}& \, N^3
\left[
-\frac{21758}{81} 
+\frac{16000}{27}  \sbz \zeta_{3}
-\frac{416}{3}  \sbz \zeta_{4}
-\frac{1280}{3}  \sbz \zeta_{5}
\right]
{+} \, N^4
\left[
\frac{856}{243} 
+\frac{128}{27}  \sbz \zeta_{3}
\right]
\Biggr)
\label{beMS5l}
{}.
\eea

Traditionally in calculations of leptonic($\ell$) anomalies $a_\ell$
the classical OS scheme is employed. In this subtraction scheme the
QED is parametrized with two variables, namely, the fine structure
constant $\al\equiv \al^{OS}$, equal to  the  invariant charge \re{eq:InvCharge} at zero momentum transfer,
and the pole mass $M$ of the fermion.
The corresponding normalization condition for the OS polarization function
$\Pi(-q^2/M^2,\al) \equiv \Pi^{OS}(-q^2/M^2,\al)$ is:
\begin{equation} 
\label{eq:OScharge}
\Pi(-q^2=0,M,\alpha) \equiv 0
{}.
\end{equation} 
The {\em asymptotic} polarization function $\Pi^{asymp}(Q^2/M^2,\al)$
is obtained from $\Pi(Q^2/M^2,\al)$ by discarding in every order of
perturbation theory all power suppressed terms in the limit of $M \to
0$. Finally, according  to Ref.~\cite{Lautrup:1974ic} 
one  obtains the asymptotic part of the muon anomaly (related to the vacuum polarization diagrams)
through  the relation:
\begin{equation}
\label{eq:amuasymp}
  a^{\mbox{\scriptsize asymp}}_{\mu}={\alpha \over \pi}\int_{0}^{1}\!dx\*(1-x)\*
  \left[d^{\mbox{\scriptsize
        asymp}}_{R}\left({x^2\over1-x}{M_\mu^2\over
        M_e^2},\alpha\right)-1\right] 
,
\end{equation}
where 
\begin{equation}
\label{eq:dR}
d^{\mbox{\scriptsize asymp}}_R(Q^2/M^2,\alpha)={1\over 1+\Pi^{\mbox{\scriptsize asymp}}(Q^2/M^2,\alpha)}.
\end{equation}

\section{The polarization function at four loops\label{sec:VacPol}}
The asymptotic limit $\ovl{m}\to0$
requires the computation of massless propagators, whereas the limit
$q^2\to0$ leads to the evaluation of massive tadpole diagrams. The
perturbative expansion of the vacuum polarization function in the fine structure
constant in the two cases is conveniently defined by:
\begin{equation}
\label{eq:expm0}
\ovl{\Pi}(Q^2,\ovl{m}^2 =0,\ovl{\alpha}) = 
\sum_{i} \ovl{\Pi}^{(i)}(\lmuQ)\left(\frac{\ovl{ \alpha}(\mu)}{\pi} \right)^i
\end{equation}
and 
\begin{equation}
\label{eq:expQ0}
\ovl{\Pi}(Q^2 =0 ,\ovl{m}^2,\ovl{\alpha}) = 
\sum_{i} \ovl{\Pi}^{(i)}(q=0,\ell_{{\mu} m})\left(\frac{\ovl{ \alpha}(\mu)}{\pi} \right)^i
{},
\end{equation}
where   $\ell_{{\mu}Q}=\ln(\mu^2/Q^2)$ and   $\ell_{{\mu} m}=\ln(\mu^2/\ovl{m}^2)$.
In both cases the computation has been performed with
{\tt{FORM}}~\cite{Vermaseren:2000nd,Vermaseren:2002rp,Tentyukov:2006ys}
based programs. The four-loop QED diagrams contributing to the electron
polarization function have been generated with the program
{\tt{QGRAF}}\cite{Nogueira:1991ex}.

The  four-loop massless propagators can be  reduced to 28 master integrals. The reduction of 
four-loop massless propagators  has been done by  evaluating  
sufficiently many terms of the $1/D$ expansion \cite{Baikov:2005nv} of
the corresponding coefficient functions \cite{Baikov:1996rk}. The master integrals are known analytically from
\cite{Baikov:2010hf,Lee:2011jt}.

In the low $Q^2$ limit all appearing tadpole diagrams
have been reduced to master integrals with the help of Laporta's
algorithm~\cite{Laporta:1996mq,Laporta:2001dd}. The arising polynomials
in the space-time dimension $d=4-2\*\vep$ have been simplified with the
program {\tt{FERMAT}}~\cite{Lewis}. The remaining master integrals are
known analytically to sufficient high order in $\vep$ and have been
taken from
Refs.~\cite{Laporta:2002pg,Chetyrkin:2004fq,Kniehl:2005yc,Schroder:2005va,Schroder:2005db,Chetyrkin:2006dh,Bejdakic:2006vg,Kniehl:2006bf,Kniehl:2006bg}.

For the large
$Q^2$ limit our result, renormalized completely in the $\MSbar$ scheme, reads: 
\begin{eqnarray}
\ovl{\Pi}^{(1)}(\ell_{\mu Q} ) &=&  
 N 
\left[
\frac{5}{9} 
+\frac{1}{3} \ell_{\mu Q}
\right]
{},
\label{PiMS1}
\\
\ovl{\Pi}^{(2)}(\ell_{\mu Q} ) &=&  
 \, N 
\left[
\frac{55}{48} 
-  \sbz \zeta_{3}
+\frac{1}{4} \ell_{\mu Q}
\right]
{},
\label{PiMS2}
\\
\ovl{\Pi}^{(3)}(\ell_{\mu Q} ) &=&  
 N 
\left[
-\frac{143}{288} 
-\frac{37}{24}  \sbz \zeta_{3}
+\frac{5}{2}  \sbz \zeta_{5}
-\frac{1}{32} \ell_{\mu Q}
\right]
\nonumber\\
&{+}& \, N^2
\left[
-\frac{3701}{2592} 
+\frac{19}{18}  \sbz \zeta_{3}
-\frac{11}{24} \ell_{\mu Q}
+\frac{1}{3}  \sbz \zeta_{3}\ell_{\mu Q}
-\frac{1}{24} \ell_{\mu Q}^2
\right]
{},
\label{PiMS3}
\\
\ovl{\Pi}^{(4)}(\ell_{\mu Q} ) &=& 
 \, N 
\left[
-\frac{31}{768} 
+\frac{13}{32}  \sbz \zeta_{3}
+\frac{245}{32}  \sbz \zeta_{5}
-\frac{35}{4}  \,\zeta_{7}
-\frac{23}{128} \ell_{\mu Q}
\right]
\nonumber\\
&{+}& \, N^2
\left[
-\frac{7505}{41472} 
+\frac{11}{8640} \,\pi^4
+\frac{1553}{216}  \sbz \zeta_{3}
-  \,\zeta_3^2
-\frac{125}{18}  \sbz \zeta_{5}
-\frac{29}{192} \ell_{\mu Q}
\BreakI
\phantom{+ \, N^2}
+\frac{19}{12}  \sbz \zeta_{3}\ell_{\mu Q}
-\frac{5}{3}  \sbz \zeta_{5}\ell_{\mu Q}
-\frac{1}{48} \ell_{\mu Q}^2
\right]
\nonumber\\
&{+}&\,N^2\,{\bf si}\,
\left[
\frac{431}{432} 
-\frac{1}{360} \,\pi^4
-\frac{21}{16}  \sbz \zeta_{3}
-\frac{2}{3}  \,\zeta_3^2
+\frac{5}{4}  \sbz \zeta_{5}
+\frac{11}{36} \ell_{\mu Q}
-\frac{2}{3}  \sbz \zeta_{3}\ell_{\mu Q}
\right]
\nonumber\\
&{+}& \, N^3
\left[
\frac{196513}{93312} 
-\frac{809}{648}  \sbz \zeta_{3}
-\frac{5}{9}  \sbz \zeta_{5}
+\frac{151}{162} \ell_{\mu Q}
-\frac{19}{27}  \sbz \zeta_{3}\ell_{\mu Q}
+\frac{11}{72} \ell_{\mu Q}^2
\BreakI
\phantom{+ \, N^3}
-\frac{1}{9}  \sbz \zeta_{3}\ell_{\mu Q}^2
+\frac{1}{108} \ell_{\mu Q}^3
\right]
{},
\label{PiMS4}
\end{eqnarray}
%
%
%
with $\zeta_n$ being the Riemann zeta-function defined by:
\begin{equation}
\label{eq:zeta}
\zeta_n=\sum_{k=1}^{\infty}{1\over k^n}
\end{equation}
and $\ell_{\mu Q} = \ln(\frac{\mu^2}{Q^2})$.
The symbol ${\bf{si}}$ is here and in the following equal one and serves
only as a separator in order to display the singlet contributions,
originating from diagrams of the type as shown in
Fig.~\ref{fig:VPclassification}(s), separately. As stated above,
    no logarithms of the type $\ln(\ovl{m}^2)$ appear in this
    result. However, in the low $Q^2=0$ case the renormalized vacuum
polarization function does contain logarithms of the type $\ell_{\mu m} =
\ln({\mu^2\over \ovl{m}^2})$. In this limit our analytical result,
renormalized in the $\MSbar$ scheme, reads:
\begin{eqnarray}
\ovl{\Pi}^{(1)} (q=0,\ell_{\mu m}) &=&
 N 
\left[
\frac{1}{3} \,\ell_{\mu m}\,
\right]
{},
\label{Pi0MS1}
\\
\ovl{\Pi}^{(2)} (q=0,\ell_{\mu m}) &=&
 N 
\left[
\frac{13}{48} 
-\frac{1}{4} \,\ell_{\mu m}\,
\right]
{},
\label{Pi0MS2}
\\
\ovl{\Pi}^{(3)} (q=0,\ell_{\mu m}) &=&
 N 
\left[
\frac{97}{288} 
-\frac{95}{192}  \sbz \zeta_{3}
+\frac{9}{32} \,\ell_{\mu m}\,
\right]
\nonumber\\
&{+}& \, N^2
\left[
-\frac{103}{1296} 
+\frac{7}{64}  \sbz \zeta_{3}
-\frac{1}{36} \,\ell_{\mu m}\,
+\frac{1}{24} \,\ell_{\mu m}^2
\right]
{},
\label{Pi0MS3}
\eea
\bea
\lefteqn{\ovl{\Pi}^{(4)} (q=0,\ell_{\mu m}) = }
\nonumber
\\
&{}&
 N 
\left[
-\frac{37441}{34560} 
+\frac{58001}{129600} \,\pi^4
-\frac{7549}{320}  \sbz \zeta_{3}
+\frac{3429}{160}  \sbz \zeta_{5}
-\frac{157}{128} \,\ell_{\mu m}\,
-\frac{106}{675} \,\pi^4 \logtwo \, 
\BreakI
+\frac{1919}{1080} \,\pi^2 \Log{2}{2}\,
-\frac{32}{135} \,\pi^2 \Log{2}{3}\,
-\frac{1919}{1080}  \Log{2}{4}\,
+\frac{32}{225}  \Log{2}{5}\,
-\frac{1919}{45}  \,a_4
-\frac{256}{15}  \,a_5
\right]
\nonumber\\
&{+}& \, N^2
\left[
-\frac{2261597}{1036800} 
+\frac{29737}{129600} \,\pi^4
-\frac{123149}{10800}  \sbz \zeta_{3}
-\frac{41}{576} \,\ell_{\mu m}\,
+\frac{13}{96}  \sbz \zeta_{3}\,\ell_{\mu m}\,
-\frac{1}{8} \,\ell_{\mu m}^2
\BreakI
\phantom{+ \, N^2}
+\frac{437}{540} \,\pi^2 \Log{2}{2}\,
-\frac{437}{540}  \Log{2}{4}\,
-\frac{874}{45}  \,a_4
\right]
\nonumber\\
&{+}&\,N^2\,{\bf si}\,
\left[
\frac{2411}{5040} 
+\frac{2189}{17280} \,\pi^4
-\frac{6779}{1120}  \sbz \zeta_{3}
-\frac{5}{12}  \sbz \zeta_{5}
+\frac{11}{36} \,\ell_{\mu m}\,
-\frac{2}{3}  \sbz \zeta_{3}\,\ell_{\mu m}\,
\BreakI
\phantom{+\,N^2\,{\bf si}\,}
+\frac{73}{144} \,\pi^2 \Log{2}{2}\,
-\frac{73}{144}  \Log{2}{4}\,
-\frac{73}{6}  \,a_4
\right]
\nonumber\\
&{+}& \, N^3
\left[
\frac{610843}{3265920} 
-\frac{661}{3780}  \sbz \zeta_{3}
+\frac{113}{1296} \,\ell_{\mu m}\,
-\frac{7}{96}  \sbz \zeta_{3}\,\ell_{\mu m}\,
+\frac{1}{108} \,\ell_{\mu m}^2
-\frac{1}{108} \,\ell_{\mu m}^3
\right]
{},
\label{Pi0MS4}
\end{eqnarray}
%
%
%
where the polylogarithm function $\mbox{Li}_n(1/2)$ is defined by:
\begin{equation}
\label{eq:a}
\Pa{n}=\mbox{Li}_n(1/2)=\sum_{k=1}^{\infty}{1\over2^{k}k^{n}}\,.
\end{equation}

In the next step we convert the results of
Eqs.~(\ref{PiMS1})-(\ref{PiMS4}) from the $\MSbar$ into the on-shell scheme in
order to obtain the asymptotic limit of the polarization function
$\Pi^{\mbox{\scriptsize asymp}}(Q^2 ,M^2,\alpha)$ in OS scheme. 
The scheme independence of the invariant charge directly leads to the relation:
\beq
\Pi^{\asy}(Q^2/M^2,\al) = \frac{\al}{\oal}
\Biggl(
1 + \ovl{\Pi}^{\asy}(Q^2/\mu^2,\oal)
\Biggr) -1
\label{MS:OS}
{},
\eeq 
where it is understood that the running $\oal$ on the r.h.s. is converted 
into the on-shell coupling constant $\al$. The corresponding conversion relation is given in the 
next Section~\ref{sec:alphaconversion}.

In analogy to Eqs.~(\ref{eq:expm0}) and (\ref{eq:expQ0}) the expansion in
the on-shell fine structure constant $\alpha$ of $\Pi^{\mbox{\scriptsize
    asymp}}(Q^2 ,M^2,\alpha)$ is defined by:
\begin{equation}
\label{eq:expPiOS}
\Pi^{\mbox{\scriptsize asymp}}(Q^2/M^2,\alpha) = 
\sum_{i} {\Pi}^{\mbox{\scriptsize asymp},(i)}(\ellMQ)\left(\frac{\alpha}{\pi}\right)^i
{}.
\end{equation}
The asymptotic limit of the polarization function in the OS scheme is given by
($\ell_{M Q} = \ln(\frac{M^2}{Q^2})$):
\begin{eqnarray}
\Pi^{\mbox{\scriptsize asymp},(1)}(\ell_{M Q}) &=&
 N 
\left[
\frac{5}{9} 
+\frac{1}{3} \ell_{MQ}
\right]
{},
\label{PiOs1}
\\
\Pi^{\mbox{\scriptsize asymp},(2)}(\ell_{M Q}) &=&  
 N 
\left[
\frac{5}{24} 
-  \sbz \zeta_{3}
+\frac{1}{4} \ell_{MQ}
\right]
{},
\label{PiOs2}
\\
\Pi^{\mbox{\scriptsize asymp},(3)}(\ell_{M Q}) &=& 
 \, N 
\left[
-\frac{121}{192} 
-\frac{5}{24} \,\pi^2
-\frac{99}{64}  \sbz \zeta_{3}
+\frac{5}{2}  \sbz \zeta_{5}
-\frac{1}{32} \ell_{MQ}
+\frac{1}{3} \,\pi^2 \logtwo \, 
\right]
\nonumber\\
&{+}& \, N^2
\left[
-\frac{307}{864} 
-\frac{1}{9} \,\pi^2
+\frac{545}{576}  \sbz \zeta_{3}
-\frac{11}{24} \ell_{MQ}
+\frac{1}{3}  \sbz \zeta_{3}\ell_{MQ}
-\frac{1}{24} \ell_{MQ}^2
\right]
{},
\label{PiOs3}
\eea
\bea
&{}&
\Pi^{\mbox{\scriptsize asymp},(4)}(\ell_{M Q}) =   
 N 
\left[
-\frac{71189}{34560} 
-\frac{157}{72} \,\pi^2
-\frac{59801}{129600} \,\pi^4
+\frac{6559}{320}  \sbz \zeta_{3}
-\frac{1}{24} \,\pi^2 \sbz \zeta_{3}
-\frac{1603}{120}  \sbz \zeta_{5}
\BreakI
\phantom{+ \, N }
-\frac{35}{4}  \,\zeta_{7}
-\frac{23}{128} \ell_{MQ}
+\frac{59}{12} \,\pi^2 \logtwo \, 
+\frac{106}{675} \,\pi^4 \logtwo \, 
-\frac{1559}{1080} \,\pi^2 \Log{2}{2}\,
+\frac{32}{135} \,\pi^2 \Log{2}{3}\,
\BreakI
\phantom{+ \, N }
+\frac{1559}{1080}  \Log{2}{4}\,
-\frac{32}{225}  \Log{2}{5}\,
+\frac{1559}{45}  \,a_4
+\frac{256}{15}  \,a_5
\right]
\nonumber\\
&{+}& \, N^2
\left[
\frac{3361}{900} 
-\frac{179}{324} \,\pi^2
-\frac{2161}{10800} \,\pi^4
+\frac{29129}{1800}  \sbz \zeta_{3}
-  \,\zeta_3^2
-\frac{125}{18}  \sbz \zeta_{5}
\BreakI
\phantom{+ \, N^2}
+\frac{1}{12} \ell_{MQ}
+\frac{19}{12}  \sbz \zeta_{3}\ell_{MQ}
-\frac{5}{3}  \sbz \zeta_{5}\ell_{MQ}
-\frac{1}{48} \ell_{MQ}^2
+\frac{16}{27} \,\pi^2 \logtwo \, 
-\frac{53}{60} \,\pi^2 \Log{2}{2}\,
\BreakI
\phantom{+ \, N^2}
+\frac{53}{60}  \Log{2}{4}\,
+\frac{106}{5}  \,a_4
\right]
\nonumber\\
&{+}&\,N^2\,{\bf si}\,
\left[
\frac{1963}{3780} 
-\frac{2237}{17280} \,\pi^4
+\frac{5309}{1120}  \sbz \zeta_{3}
-\frac{2}{3}  \,\zeta_3^2
+\frac{5}{3}  \sbz \zeta_{5}
+\frac{11}{36} \ell_{MQ}
\BreakI
\phantom{+\,N^2\,{\bf si}\,}
-\frac{2}{3}  \sbz \zeta_{3}\ell_{MQ}
-\frac{73}{144} \,\pi^2 \Log{2}{2}\,
+\frac{73}{144}  \Log{2}{4}\,
+\frac{73}{6}  \,a_4
\right]
\nonumber\\
&{+}& \, N^3
\left[
\frac{75259}{68040} 
+\frac{8}{405} \,\pi^2
-\frac{15109}{22680}  \sbz \zeta_{3}
-\frac{5}{9}  \sbz \zeta_{5}
+\frac{151}{162} \ell_{MQ}
-\frac{19}{27}  \sbz \zeta_{3}\ell_{MQ}
\BreakI
\phantom{+ \, N^3}
+\frac{11}{72} \ell_{MQ}^2
-\frac{1}{9}  \sbz \zeta_{3}\ell_{MQ}^2
+\frac{1}{108} \ell_{MQ}^3
\right]
{}.
\label{PiOs4}
\end{eqnarray}
The results displayed in Eqs.~(\ref{PiOs1})-(\ref{PiOs3}) (as well as
those in Eqs.~(\ref{PiMS1})-(\ref{PiMS3}), (\ref{Pi0MS1})-(\ref{Pi0MS3}))
are known since long (see, e.g. \cite{Broadhurst:1992za} and
references therein).  Note that all terms proportional to $\ell_{MQ}$ in
Eq.~(\ref{PiOs4}) follow in an easy way from renormalization group
arguments as explained in \cite{Lautrup:1974ic}. In addition, in the
process of constructing $d_R^{\asy}$ one should invert the power series
for $(1+ \Pi^{\asy})$, which also produces
{childishly-easy-to-compute} factorizable fourth order contributions
to $d_R^{\asy}$ like $\alpha^4 \, \Pi^{\asy,(2)}\,\Pi^{\asy,(2)}$
and so on (see e.g.  \cite{Kataev:2006gx} and references therein).
However, to include power suppressed terms of order $(M/Q)^n$ to
$\Pi^{\mathrm{OS}}$ is much less trivial even for the factorizable
contributions \cite{Czarnecki:1998rc,Aguilar:2008qj}.  In fact, the
second work also provides power suppressed contributions to the muon
anomaly from factorizable diagrams of type I(a) (see
Fig.~\ref{fig:MuonClasses}) for the cases when leptons circulated in
closed loops could be not just electrons but also muons and
tau-leptons in different combinations.

\section{The QED charge renormalized in $\MSbar$ and on-shell scheme\label{sec:alphaconversion}}
Let us define the conversion factor $C_{\alpha\ovl{\alpha}}$, which
converts the fine structure constant $\ovl{\alpha}$ in $\MSbar$ scheme
into $\alpha$ in OS scheme: $\alpha = C_{\alpha\ovl{\alpha}}\,
\ovl{\alpha}$. The conversion factor $C_{\alpha\ovl{\alpha}}$ has a
perturbative expansion in $\ovl{\alpha}$ defined by:
\begin{equation}
\label{eq:expCaabar}
C_{\alpha\ovl{\alpha}} = 1+ \sum_{i \ge 1}C_{\alpha\ovl{\alpha}}^{(i)} 
\left(\frac{\ovl{ \alpha}(\mu)}{\pi} \right)^i.
\end{equation}
The expansion coefficients $C_{\alpha\ovl{\alpha}}^{(i)}$ can be
obtained by evaluating Eq.~(\ref{MS:OS}) at $Q^2=0$ and using Eq.~(\ref{eq:OScharge}). On the r.h.s.
of this equation we insert the results of
Eqs.~(\ref{eq:expQ0}), (\ref{Pi0MS1})-(\ref{Pi0MS4}) and expand in
$\ovl{\alpha}$ up to four-loop order. 
In addition, one should use  the relation to convert
the $\MSbar$-mass $\ovl{m}(\mu)$ to the on-shell mass $M$, which is known
from  Refs.~\cite{Chetyrkin:1999qi,Melnikov:2000qh} to three-loop order. 
For the case of QED it is given in Appendix~\ref{app:MassMSbarOnShell}.

The different orders $C_{\alpha\ovl{\alpha}}^{(i)}$ read:
\begin{eqnarray}
\lefteqn{C_{\alpha\ovl{\alpha}}^{(1)}(\ell_{\mu M})= } 
\nonumber\\
&{+}& \, N 
\left[
-\frac{1}{3} \,\ell_{\mu M}\,
\right]
{},
\label{bsOSfromMS1}
\end{eqnarray}
\begin{eqnarray}
\lefteqn{C_{\alpha\ovl{\alpha}}^{(2)}(\ell_{\mu M})= } 
\nonumber\\
&{+}& \, N 
\left[
-\frac{15}{16} 
-\frac{1}{4} \,\ell_{\mu M}\,
\right]
\nonumber\\
&{+}& \, N^2
\left[
\frac{1}{9} \,\ell_{\mu M}^2
\right]
{},
\label{bsOSfromMS2}
\end{eqnarray}
\begin{eqnarray}
\lefteqn{C_{\alpha\ovl{\alpha}}^{(3)}(\ell_{\mu M})= } 
\nonumber\\
&{+}& \, N 
\left[
-\frac{77}{576} 
-\frac{5}{24} \,\pi^2
-\frac{1}{192}  \sbz \zeta_{3}
+\frac{1}{32} \,\ell_{\mu M}\,
+\frac{1}{3} \,\pi^2 \logtwo \, 
\right]
\nonumber\\
&{+}& \, N^2
\left[
\frac{695}{648} 
-\frac{1}{9} \,\pi^2
-\frac{7}{64}  \sbz \zeta_{3}
+\frac{73}{72} \,\ell_{\mu M}\,
+\frac{5}{24} \,\ell_{\mu M}^2
\right]
\nonumber\\
&{+}& \, N^3
\left[
-\frac{1}{27} \,\ell_{\mu M}^3
\right]
{},
\label{bsOSfromMS3}
\end{eqnarray}
\begin{eqnarray}
\lefteqn{C_{\alpha\ovl{\alpha}}^{(4)}(\ell_{\mu M})= } 
\nonumber\\
&{+}& \, N 
\left[
-\frac{34897}{17280} 
-\frac{157}{72} \,\pi^2
-\frac{59801}{129600} \,\pi^4
+\frac{6429}{320}  \sbz \zeta_{3}
-\frac{1}{24} \,\pi^2 \sbz \zeta_{3}
-\frac{10087}{480}  \sbz \zeta_{5}
\BreakI
\phantom{+ \, N }
+\frac{23}{128} \,\ell_{\mu M}\,
+\frac{59}{12} \,\pi^2 \logtwo \, 
+\frac{106}{675} \,\pi^4 \logtwo \, 
-\frac{1559}{1080} \,\pi^2 \Log{2}{2}\,
\BreakI
\phantom{+ \, N }
+\frac{32}{135} \,\pi^2 \Log{2}{3}\,
+\frac{1559}{1080}  \Log{2}{4}\,
-\frac{32}{225}  \Log{2}{5}\,
+\frac{1559}{45}  \,a_4
+\frac{256}{15}  \,a_5
\right]
\nonumber\\
&{+}& \, N^2
\left[
\frac{4768247}{1036800} 
-\frac{179}{324} \,\pi^2
-\frac{8699}{43200} \,\pi^4
+\frac{107249}{10800}  \sbz \zeta_{3}
+\frac{1861}{1728} \,\ell_{\mu M}\,
+\frac{5}{18} \,\pi^2\,\ell_{\mu M}\,
\BreakI
\phantom{+ \, N^2}
-\frac{43}{144}  \sbz \zeta_{3}\,\ell_{\mu M}\,
+\frac{1}{16} \,\ell_{\mu M}^2
+\frac{16}{27} \,\pi^2 \logtwo \, 
-\frac{4}{9} \,\pi^2\,\ell_{\mu M}\, \logtwo \, 
-\frac{53}{60} \,\pi^2 \Log{2}{2}\,
\BreakI
\phantom{+ \, N^2}
+\frac{53}{60}  \Log{2}{4}\,
+\frac{106}{5}  \,a_4
\right]
\nonumber\\
&{+}&\,N^2\,{\bf si}\,
\left[
-\frac{2411}{5040} 
-\frac{2189}{17280} \,\pi^4
+\frac{6779}{1120}  \sbz \zeta_{3}
+\frac{5}{12}  \sbz \zeta_{5}
-\frac{11}{36} \,\ell_{\mu M}\,
+\frac{2}{3}  \sbz \zeta_{3}\,\ell_{\mu M}\,
\BreakI
\phantom{+\,N^2\,{\bf si}\,}
-\frac{73}{144} \,\pi^2 \Log{2}{2}\,
+\frac{73}{144}  \Log{2}{4}\,
+\frac{73}{6}  \,a_4
\right]
\nonumber\\
&{+}& \, N^3
\left[
-\frac{3265523}{3265920} 
+\frac{8}{405} \,\pi^2
+\frac{2201}{3780}  \sbz \zeta_{3}
-\frac{5483}{3888} \,\ell_{\mu M}\,
+\frac{4}{27} \,\pi^2\,\ell_{\mu M}\,
+\frac{7}{48}  \sbz \zeta_{3}\,\ell_{\mu M}\,
\BreakI
\phantom{+ \, N^3}
-\frac{101}{144} \,\ell_{\mu M}^2
-\frac{13}{108} \,\ell_{\mu M}^3
\right]
\nonumber\\
&{+}& \, N^4
\left[
\frac{1}{81} \,\ell_{\mu M}^4
\right]
{},
\label{bsOSfromMS4}
\end{eqnarray}
%
%
with $\ell_{{\mu}M}=\ln(\mu^2/M^2)$.
On the other hand the inverse conversion factor
$C_{\ovl{\alpha}{\alpha}}$, which allows a conversion from the $\MSbar$ to
the on-shell scheme $\ovl{\alpha} =C_{\ovl{\alpha}{\alpha}}\, \alpha$ is
useful as well. In analogy to Eq.~(\ref{eq:expCaabar}) the perturbative
expansion in $\alpha$ of $C_{\ovl{\alpha}{\alpha}}$ is defined as:
\begin{equation} 
C_{\ovl{\alpha}\alpha} = 1+ \sum_{i \ge 1}C_{\ovl{\alpha}\alpha}^{(i)} 
\left(\frac{{ \alpha}}{\pi} \right)^i.
\end{equation}
Proceeding like in the previous case and inverting
the series obtained with the help of Eq.~(\ref{MS:OS}), we find:
\begin{eqnarray}
\lefteqn{C_{\ovl{\alpha}\alpha}^{(1)}(\ell_{\mu M})=  } 
\nonumber\\
&{+}& \, N 
\left[
\frac{1}{3} \,\ell_{\mu M}\,
\right]
{},
\label{bsMSfromOS1}
\end{eqnarray}
\begin{eqnarray}
\lefteqn{C_{\ovl{\alpha}\alpha}^{(2)}(\ell_{\mu M})=  } 
\nonumber\\
&{+}& \, N 
\left[
\frac{15}{16} 
+\frac{1}{4} \,\ell_{\mu M}\,
\right]
\nonumber\\
&{+}& \, N^2
\left[
\frac{1}{9} \,\ell_{\mu M}^2
\right]
{},
\label{bsMSfromOS2}
\end{eqnarray}
\begin{eqnarray}
\lefteqn{C_{\ovl{\alpha}\alpha}^{(3)}(\ell_{\mu M})=  } 
\nonumber\\
&{+}& \, N 
\left[
\frac{77}{576} 
+\frac{5}{24} \,\pi^2
+\frac{1}{192}  \sbz \zeta_{3}
-\frac{1}{32} \,\ell_{\mu M}\,
-\frac{1}{3} \,\pi^2 \logtwo \, 
\right]
\nonumber\\
&{+}& \, N^2
\left[
-\frac{695}{648} 
+\frac{1}{9} \,\pi^2
+\frac{7}{64}  \sbz \zeta_{3}
+\frac{79}{144} \,\ell_{\mu M}\,
+\frac{5}{24} \,\ell_{\mu M}^2
\right]
\nonumber\\
&{+}& \, N^3
\left[
\frac{1}{27} \,\ell_{\mu M}^3
\right]
{},
\label{bsMSfromOS3}
\end{eqnarray}
\begin{eqnarray}
\lefteqn{C_{\ovl{\alpha}\alpha}^{(4)}(\ell_{\mu M})=  } 
\nonumber\\
&{+}& \, N 
\left[
\frac{34897}{17280} 
+\frac{157}{72} \,\pi^2
+\frac{59801}{129600} \,\pi^4
-\frac{6429}{320}  \sbz \zeta_{3}
+\frac{1}{24} \,\pi^2 \sbz \zeta_{3}
+\frac{10087}{480}  \sbz \zeta_{5}
\BreakI
\phantom{+ \, N }
-\frac{23}{128} \,\ell_{\mu M}\,
-\frac{59}{12} \,\pi^2 \logtwo \, 
-\frac{106}{675} \,\pi^4 \logtwo \, 
+\frac{1559}{1080} \,\pi^2 \Log{2}{2}\,
\BreakI
\phantom{+ \, N }
-\frac{32}{135} \,\pi^2 \Log{2}{3}\,
-\frac{1559}{1080}  \Log{2}{4}\,
+\frac{32}{225}  \Log{2}{5}\,
-\frac{1559}{45}  \,a_4
-\frac{256}{15}  \,a_5
\right]
\nonumber\\
&{+}& \, N^2
\left[
-\frac{2034497}{1036800} 
+\frac{179}{324} \,\pi^2
+\frac{8699}{43200} \,\pi^4
-\frac{107249}{10800}  \sbz \zeta_{3}
+\frac{1031}{1728} \,\ell_{\mu M}\,
+\frac{5}{36} \,\pi^2\,\ell_{\mu M}\,
\BreakI
\phantom{+ \, N^2}
+\frac{89}{288}  \sbz \zeta_{3}\,\ell_{\mu M}\,
+\frac{1}{16} \,\ell_{\mu M}^2
-\frac{16}{27} \,\pi^2 \logtwo \, 
-\frac{2}{9} \,\pi^2\,\ell_{\mu M}\, \logtwo \, 
+\frac{53}{60} \,\pi^2 \Log{2}{2}\,
\BreakI
\phantom{+ \, N^2}
-\frac{53}{60}  \Log{2}{4}\,
-\frac{106}{5}  \,a_4
\right]
\nonumber\\
&{+}&\,N^2\,{\bf si}\,
\left[
\frac{2411}{5040} 
+\frac{2189}{17280} \,\pi^4
-\frac{6779}{1120}  \sbz \zeta_{3}
-\frac{5}{12}  \sbz \zeta_{5}
+\frac{11}{36} \,\ell_{\mu M}\,
-\frac{2}{3}  \sbz \zeta_{3}\,\ell_{\mu M}\,
\BreakI
\phantom{+\,N^2\,{\bf si}\,}
+\frac{73}{144} \,\pi^2 \Log{2}{2}\,
-\frac{73}{144}  \Log{2}{4}\,
-\frac{73}{6}  \,a_4
\right]
\nonumber\\
&{+}& \, N^3
\left[
\frac{3265523}{3265920} 
-\frac{8}{405} \,\pi^2
-\frac{2201}{3780}  \sbz \zeta_{3}
-\frac{2857}{3888} \,\ell_{\mu M}\,
+\frac{2}{27} \,\pi^2\,\ell_{\mu M}\,
+\frac{7}{96}  \sbz \zeta_{3}\,\ell_{\mu M}\,
\BreakI
\phantom{+ \, N^3}
+\frac{17}{72} \,\ell_{\mu M}^2
+\frac{13}{108} \,\ell_{\mu M}^3
\right]
\nonumber\\
&{+}& \, N^4
\left[
\frac{1}{81} \,\ell_{\mu M}^4
\right]
{}.
\label{bsMSfromOS4}
\end{eqnarray}

%
%
%
%
%
%
\section{The polarization function at five loops\label{sec:PolOp5}}

In this section we will use the recent calculation \cite{Baikov:2012zm}  of $\ovl{\beta}(\oal)$ --- the QED
$\beta$-function in the $\MSbar$ scheme --- at five-loop order  to find 
the {\em $Q$-dependent part} of the  five-loop  contribution to the 
asymptotic polarization function in both $\MSbar$ and OS schemes.

\subsection{ $\MSbar$ scheme}

We start by transforming  evolution equation \re{PiMS:RG:massless} for 
the $\MSbar$-renormalized  asymptotic photon polarization into the form:
\beq
\label{PiMS:RG:conv}
\mu^2\frac{\partial}{\partial \mu^2}\,  
\ovl{\Pi}^{\mbox{\scriptsize asymp}}\left(\frac{Q^{2}}{\mu^2}, \ovl{\alpha}\right)
 = \obe(\oal)\,\Biggl(
1
- \oal^2 
\frac{\partial}{\partial \oal} \, \frac{\ovl{\Pi}^{\mbox{\scriptsize asymp}}(\frac{Q^{2}}{\mu^2}, \ovl{\alpha})}{\oal}
\Biggr)   
{}.
\eeq
Now by  direct integration of the  r.h.s. of Eq.~\re{PiMS:RG:conv} one can easily construct the Q-dependent part
of  $\ovl{\Pi}^{\mbox{\scriptsize asymp}}\left(\frac{Q^{2}}{\mu^2}, \ovl{\alpha}\right)$ 
in order $\oal^5$:
\begin{eqnarray}
\lefteqn{\ovl{\Pi}^{(5)}(\ell_{\mu Q})= N\ell_{\mu Q}\Biggl\{ } 
\nonumber\\
&{}& 
\frac{4157}{6144} 
+\frac{1}{8}  \sbz \zeta_{3}
{+} \, N 
\left[
\frac{689}{1152} 
+\frac{67}{96}  \sbz \zeta_{3}
-\frac{115}{12}  \sbz \zeta_{5}
+\frac{35}{4}  \,\zeta_{7}
+\frac{13}{128} \ell_{\mu Q}
\right]
\nonumber\\
&{+}&\,N\,{\bf si}\,
\left[
-\frac{13}{12} 
-\frac{4}{3}  \sbz \zeta_{3}
+\frac{10}{3}  \sbz \zeta_{5}
\right]
\nonumber\\
&{+}& \, N^2
\left[
\frac{5713}{5184} 
-\frac{581}{72}  \sbz \zeta_{3}
+  \,\zeta_3^2
+\frac{125}{18}  \sbz \zeta_{5}
+\frac{115}{576} \ell_{\mu Q}
-\frac{7}{8}  \sbz \zeta_{3}\ell_{\mu Q}
\BreakI
\phantom{+ \, N^2}
+\frac{5}{6}  \sbz \zeta_{5}\ell_{\mu Q}
+\frac{1}{72} \ell_{\mu Q}^2
\right]
\nonumber\\
&{+}&\,N^2\,{\bf si}\,
\left[
-\frac{149}{108} 
+\frac{13}{6}  \sbz \zeta_{3}
+\frac{2}{3}  \,\zeta_3^2
-\frac{5}{3}  \sbz \zeta_{5}
-\frac{11}{72} \ell_{\mu Q}
+\frac{1}{3}  \sbz \zeta_{3}\ell_{\mu Q}
\right]
\nonumber\\
&{+}& \, N^3
\left[
-\frac{6131}{2916} 
+\frac{203}{162}  \sbz \zeta_{3}
+\frac{5}{9}  \sbz \zeta_{5}
-\frac{151}{324} \ell_{\mu Q}
+\frac{19}{54}  \sbz \zeta_{3}\ell_{\mu Q}
-\frac{11}{216} \ell_{\mu Q}^2
\BreakI
\phantom{+ \, N^3}
+\frac{1}{27}  \sbz \zeta_{3}\ell_{\mu Q}^2
-\frac{1}{432} \ell_{\mu Q}^3
\right]
\Biggr\}
{}.
\label{PiMS5}
\end{eqnarray}

\subsection{ OS scheme}

By inspecting Eq.~\re{MS:OS} relating the photon polarization function
normalized in the $\MSbar$ and OS schemes one 
observes that  the $Q$-dependent part of the ${\cal O}(\al^5)$ OS-normalized asymptotic polarization function
is  recoverable from Eq.~\re{beMS5l}, four-loop conversion relations   and five-loop asymptotic $\MSbar$ polarization
function. The  result is:
\begin{eqnarray}
\lefteqn{{\Pi}^{(5)}(\ell_{M Q})= N \ell_{M Q}\Biggl\{ \frac{4157}{6144} +\frac{1}{8} \sbz \zeta_{3} }
\nonumber\\
&{+}& \, N 
\left[
\frac{55}{96} 
+\frac{5}{96} \,\pi^2
+\frac{179}{256}  \sbz \zeta_{3}
-\frac{115}{12}  \sbz \zeta_{5}
+\frac{35}{4}  \,\zeta_{7}
+\frac{13}{128} \ell_{MQ}
-\frac{1}{12} \,\pi^2 \logtwo \, 
\right]
\nonumber\\
&{+}&\,N\,{\bf si}\,
\left[
-\frac{13}{12} 
-\frac{4}{3}  \sbz \zeta_{3}
+\frac{10}{3}  \sbz \zeta_{5}
\right]
\nonumber\\
&{+}& \, N^2
\left[
-\frac{11}{432} 
+\frac{1}{36} \,\pi^2
-\frac{17089}{2304}  \sbz \zeta_{3}
+  \,\zeta_3^2
+\frac{125}{18}  \sbz \zeta_{5}
+\frac{35}{288} \ell_{MQ}
\BreakI
\phantom{+ \, N^2}
-\frac{7}{8}  \sbz \zeta_{3}\ell_{MQ}
+\frac{5}{6}  \sbz \zeta_{5}\ell_{MQ}
+\frac{1}{72} \ell_{MQ}^2
\right]
\nonumber\\
&{+}&\,N^2\,{\bf si}\,
\left[
-\frac{149}{108} 
+\frac{13}{6}  \sbz \zeta_{3}
+\frac{2}{3}  \,\zeta_3^2
-\frac{5}{3}  \sbz \zeta_{5}
-\frac{11}{72} \ell_{MQ}
+\frac{1}{3}  \sbz \zeta_{3}\ell_{MQ}
\right]
\nonumber\\
&{+}& \, N^3
\left[
-\frac{6131}{2916} 
+\frac{203}{162}  \sbz \zeta_{3}
+\frac{5}{9}  \sbz \zeta_{5}
-\frac{151}{324} \ell_{MQ}
+\frac{19}{54}  \sbz \zeta_{3}\ell_{MQ}
-\frac{11}{216} \ell_{MQ}^2
\BreakI
\phantom{+ \, N^3}
+\frac{1}{27}  \sbz \zeta_{3}\ell_{MQ}^2
-\frac{1}{432} \ell_{MQ}^3
\right]
\Biggr\}
{}.
\label{PiOS5}
\end{eqnarray}

\section{The QED {\boldmath{$\beta$}}-function in the OS scheme at five loops \label{sec:betaOS}}
The OS  $\beta$-function, 
\beq
\beta(\al) \equiv \beta^{OS}(\alpha) \equiv \sum_{i \ge 1} \be_i\,\alp^i
{},
\label{beOS}
\eeq
describes the evolution  of the asymptotic  photon polarization function
 with respect to the change of the  mass M, namely
(see, e.g. \cite{DeRafael:1974iv}):
\beq
\label{OSrg1}
\Biggl\{
M^2\frac{\partial}{\partial M^2}
 + 
\beta(\al)\,\left( \al \frac{\partial}{\partial \al} -1 \right) 
\Biggr\}
         (1+  \Pi^{\mbox{\scriptsize asymp}}(\ellMQ,\al)) = 0
{}.
\eeq
At the four-loop level  $\beta(\al)$ is  known  since long \cite{Broadhurst:1992za}. 
One could easily  check that Eq.~\re{OSrg1}
is indeed met by our four-loop result given in Eqs.~(\ref{eq:expPiOS})
and (\ref{PiOs4}). 
  
For our purposes it is useful to rewrite Eq.~\re{OSrg1} in the form:
\beq
\label{OSrg2}
\left(
M^2\frac{\partial}{\partial M^2}
 + 
\beta(\al)\, \al 
\frac{\partial}{\partial \al} 
\right)\frac{\Pi^{\mbox{\scriptsize asymp}}(\ellMQ,\al)}{\al} = \frac{\beta(\al)}{\al}
{}.
\eeq
Eq.~\re{OSrg2}  demonstrates that  one could construct  the next $(L+1)$ order contribution to  $\beta$ 
provided  one  has  the polarization function $\Pi^{\mbox{\scriptsize asymp}}$ in L-loops (that is up to and including the terms of order $\al^L$) and its all   logarithmic (that is proportional to $\LMQ$) terms of order $\al^{L+1}$. 
Thus, we have at our disposal all ingredients  to find $\beos$ at five loops.
The result reads:
\bea
\be_1 &=&   
\frac{N}{3}
{},
\label{beOS:1}
\\
\be_2 &=&  
 \frac{N}{4}
{},
\label{beOS:2}
\\
\be_3 &=&   
-\frac{N}{32}
-\frac{7\,N^2}{18}
{},
\hspace{12cm}
\label{beOS:3}
\\
\be_4 &=& 
-\frac{23\,N}{128}
{+} \, N^2
\left[
\frac{1}{48} 
-\frac{5}{36} \,\pi^2
-\frac{35}{96}  \sbz \zeta_{3}
+\frac{2}{9} \,\pi^2 \logtwo \, 
\right]
{+} \, N^3
\left[
\frac{901}{1296} 
-\frac{2}{27} \,\pi^2
-\frac{7}{96}  \sbz \zeta_{3}
\right]
{},
\label{beOS:4}
\end{eqnarray}
\begin{eqnarray}
\be_5 &=&   
 N 
\left[
\frac{4157}{6144} 
+\frac{1}{8}  \sbz \zeta_{3}
\right]
\nonumber\\
&{+}& \, N^2
\left[
-\frac{12493}{4320} 
-\frac{643}{288} \,\pi^2
-\frac{59801}{129600} \,\pi^4
+\frac{73423}{3840}  \sbz \zeta_{3}
-\frac{1}{24} \,\pi^2 \sbz \zeta_{3}
-\frac{2203}{120}  \sbz \zeta_{5}
\BreakI
\phantom{+ \, N^2}
+5 \,\pi^2 \logtwo \, 
+\frac{106}{675} \,\pi^4 \logtwo \, 
-\frac{1559}{1080} \,\pi^2 \Log{2}{2}\,
+\frac{32}{135} \,\pi^2 \Log{2}{3}\,
\BreakI
\phantom{+ \, N^2}
+\frac{1559}{1080}  \Log{2}{4}\,
-\frac{32}{225}  \Log{2}{5}\,
+\frac{1559}{45}  \,a_4
+\frac{256}{15}  \,a_5
\right]
\nonumber\\
&{+}& \, N^3
\left[
\frac{783211}{302400} 
-\frac{47}{81} \,\pi^2
-\frac{9491}{28800} \,\pi^4
+\frac{2222237}{134400}  \sbz \zeta_{3}
+\frac{16}{27} \,\pi^2 \logtwo \, 
-\frac{1001}{720} \,\pi^2 \Log{2}{2}\,
\BreakI
\phantom{+ \, N^3}
+\frac{1001}{720}  \Log{2}{4}\,
+\frac{1001}{30}  \,a_4
\right]
\nonumber\\
&{+}& \, N^4
\left[
-\frac{203393}{204120} 
+\frac{8}{405} \,\pi^2
+\frac{493}{840}  \sbz \zeta_{3}
\right]
{}.
\label{beOS:5}
\end{eqnarray}

\section{Contributions to the muon anomaly at five loops\label{sec:muon5}}
\begin{figure}[!ht]
\begin{center}
\begin{minipage}{3cm}
\includegraphics[bb=72 375 540 720,width=3cm]{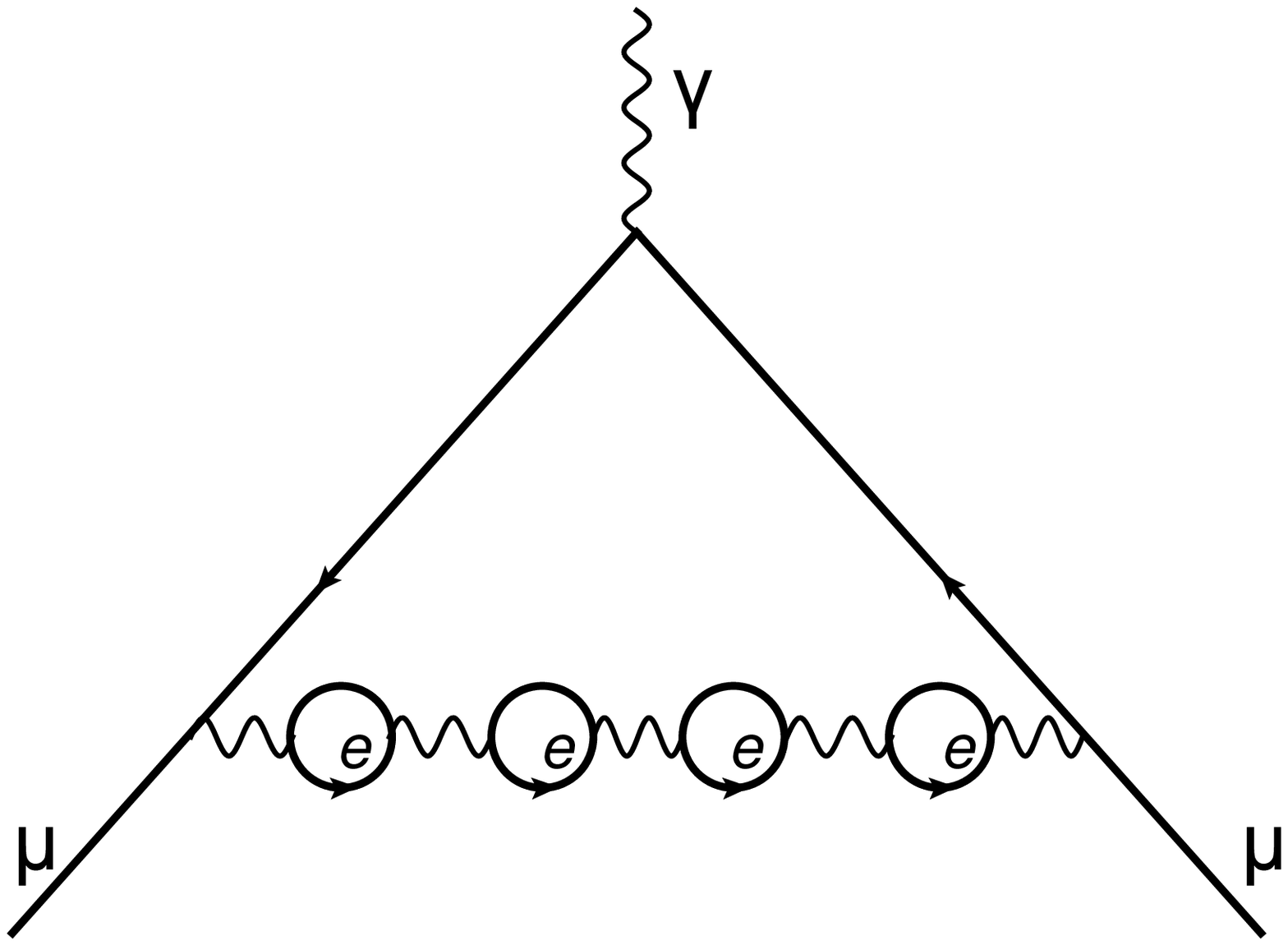}
\begin{center}
\vspace{-0.7cm}
{\tiny$I(a)$}
\end{center}
\end{minipage}
%
\begin{minipage}{3cm}
\includegraphics[bb=72 378 540 720,width=3cm]{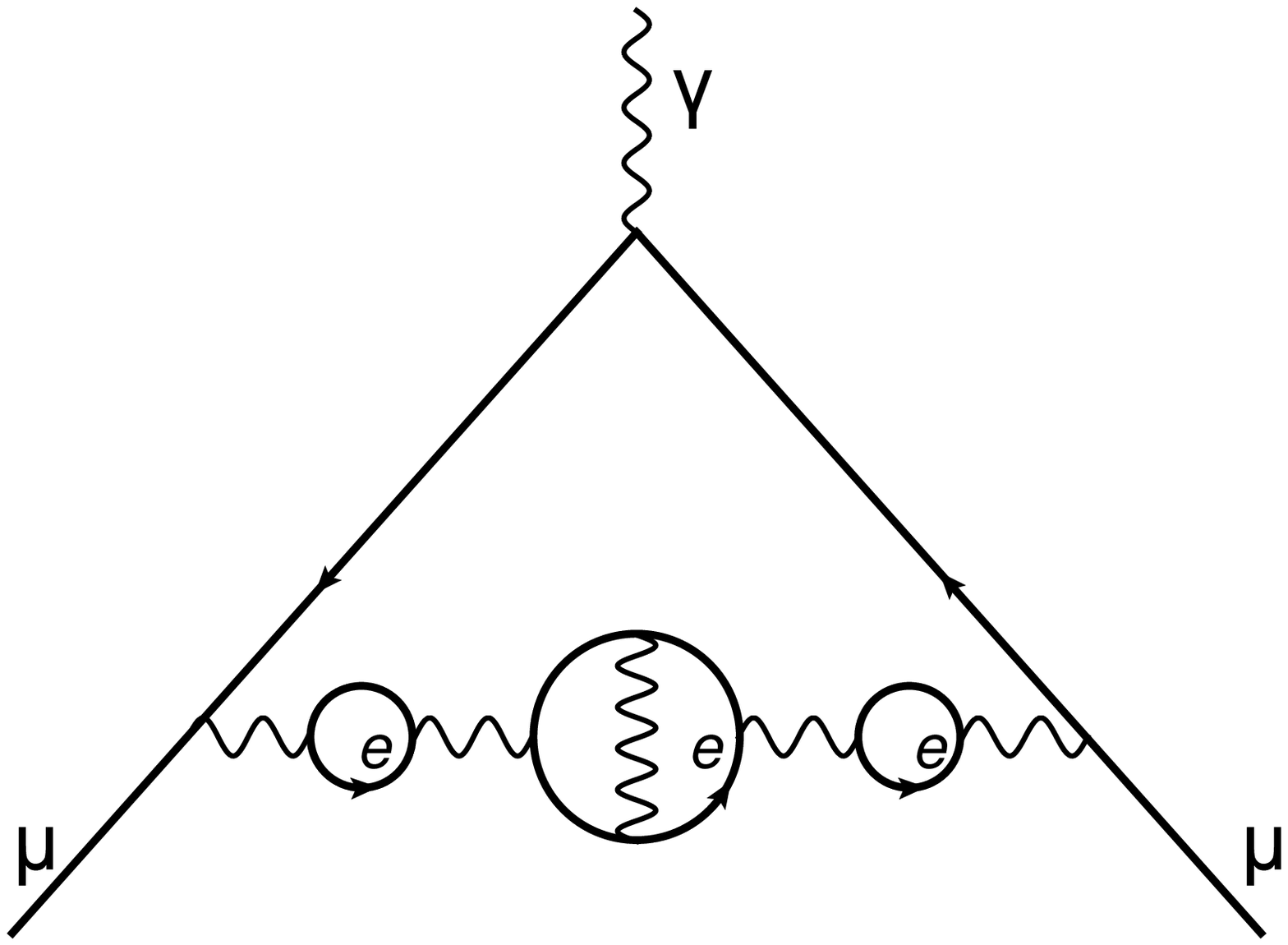}
\begin{center}
\vspace{-0.7cm}
{\tiny$I(b)$}
\end{center}
\end{minipage}
%
\begin{minipage}{3cm}
\includegraphics[bb=72 378 540 720,width=3cm]{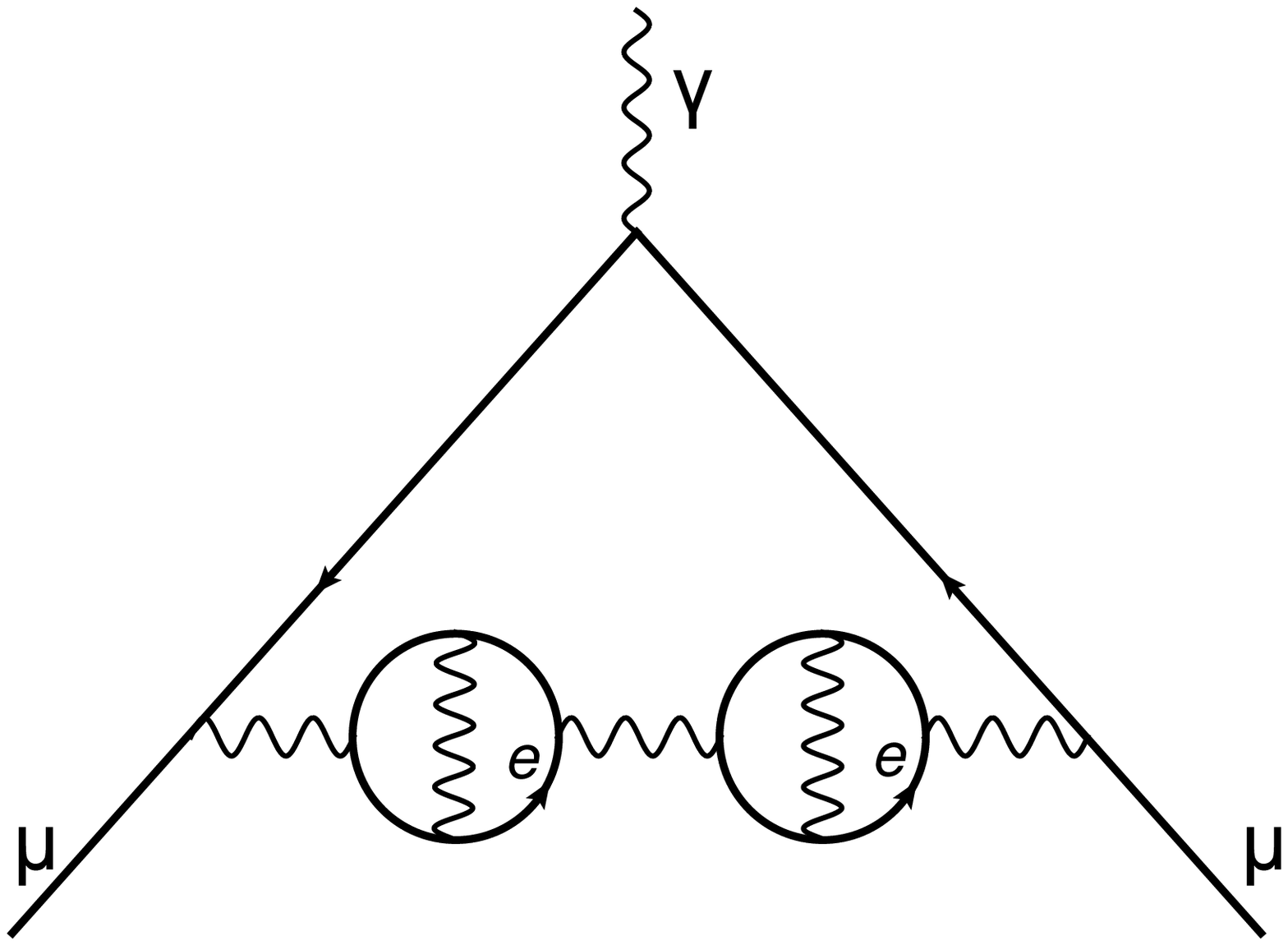}
\begin{center}
\vspace{-0.7cm}
{\tiny$I(c)$}
\end{center}
\end{minipage}
%
\begin{minipage}{3cm}
\includegraphics[bb=72 378 540 720,width=3cm]{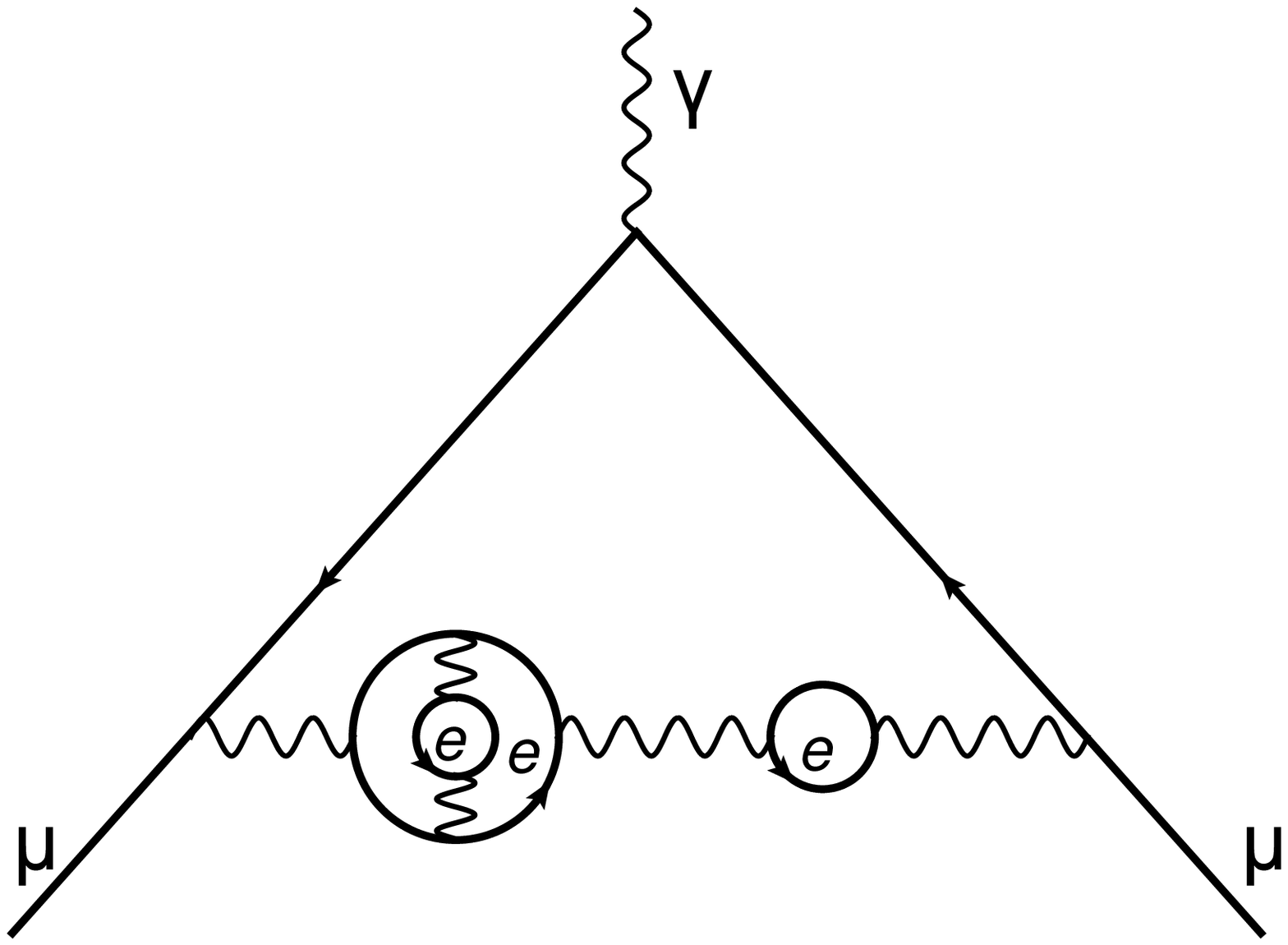}
\begin{center}
\vspace{-0.7cm}
{\tiny$I(d)$}
\end{center}
\end{minipage}
%
\begin{minipage}{3cm}
\includegraphics[bb=72 378 540 720,width=3cm]{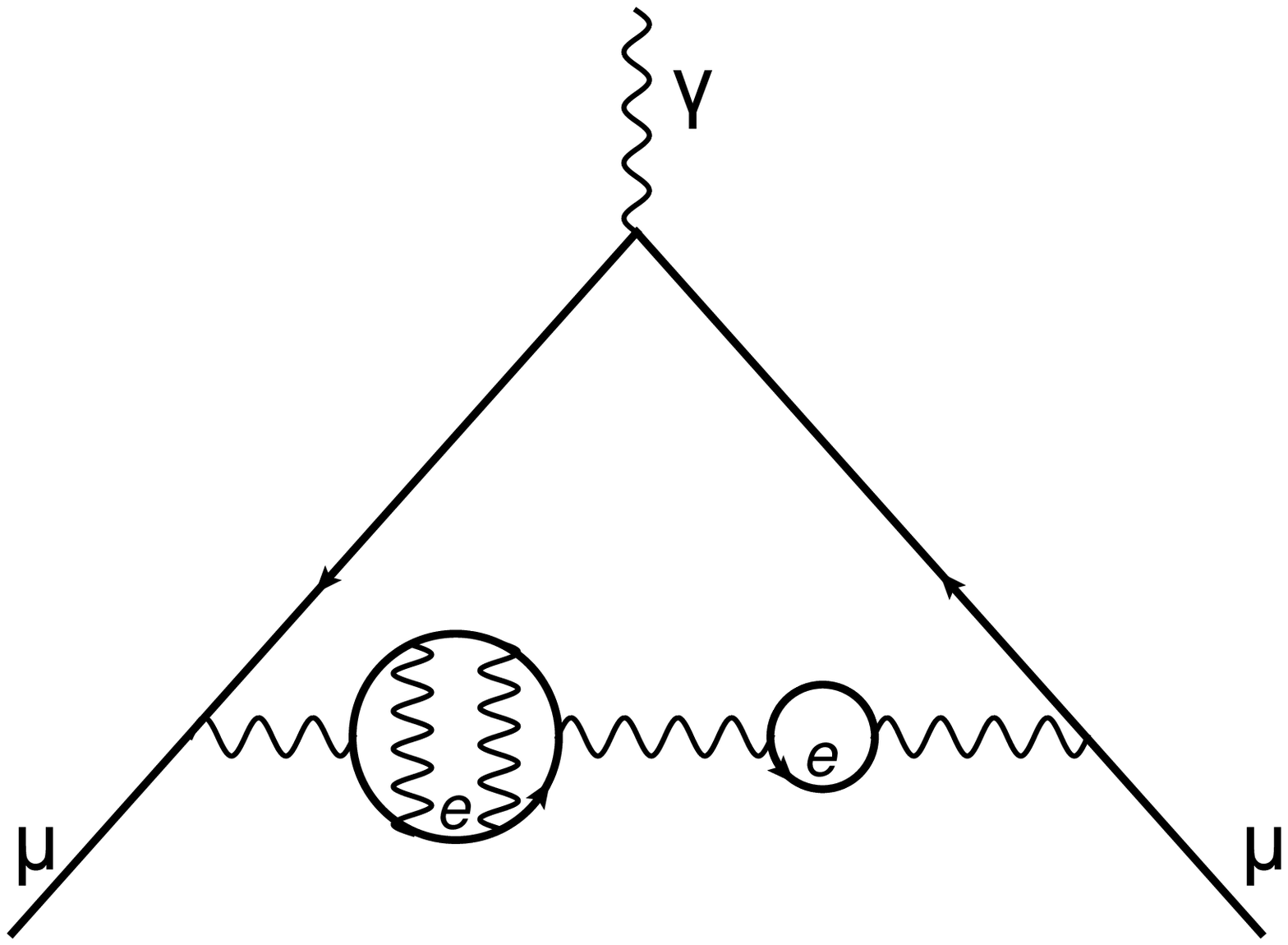}
\begin{center}
\vspace{-0.7cm}
{\tiny$I(e)$}
\end{center}
\end{minipage}\\[0.3cm]
\begin{minipage}{3cm}
\includegraphics[bb=72 378 540 720,width=3cm]{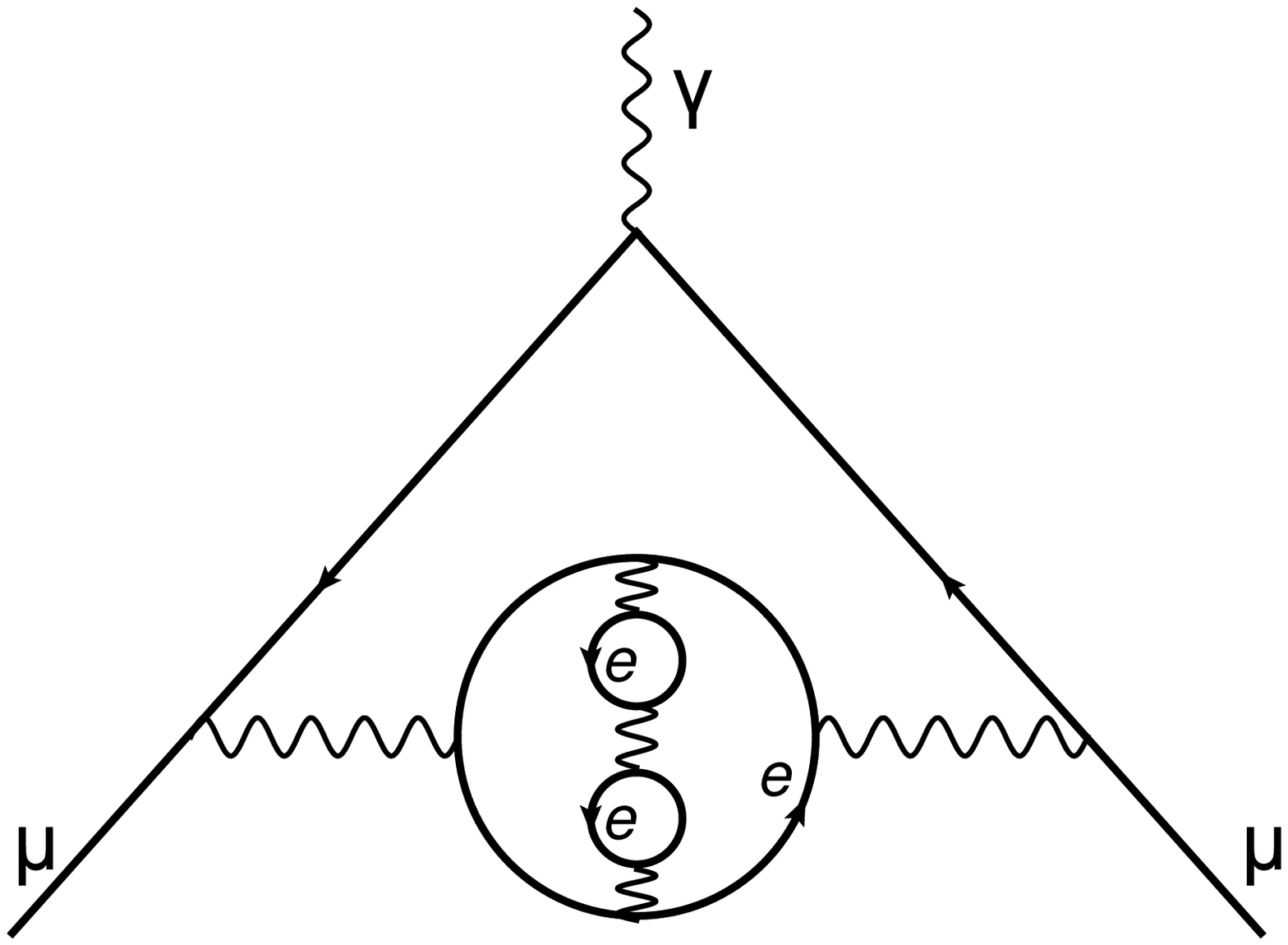}
\begin{center}
\vspace{-0.7cm}
{\tiny$I(f)$}
\end{center}
\end{minipage}
%
\begin{minipage}{3cm}
\includegraphics[bb=72 378 540 720,width=3cm]{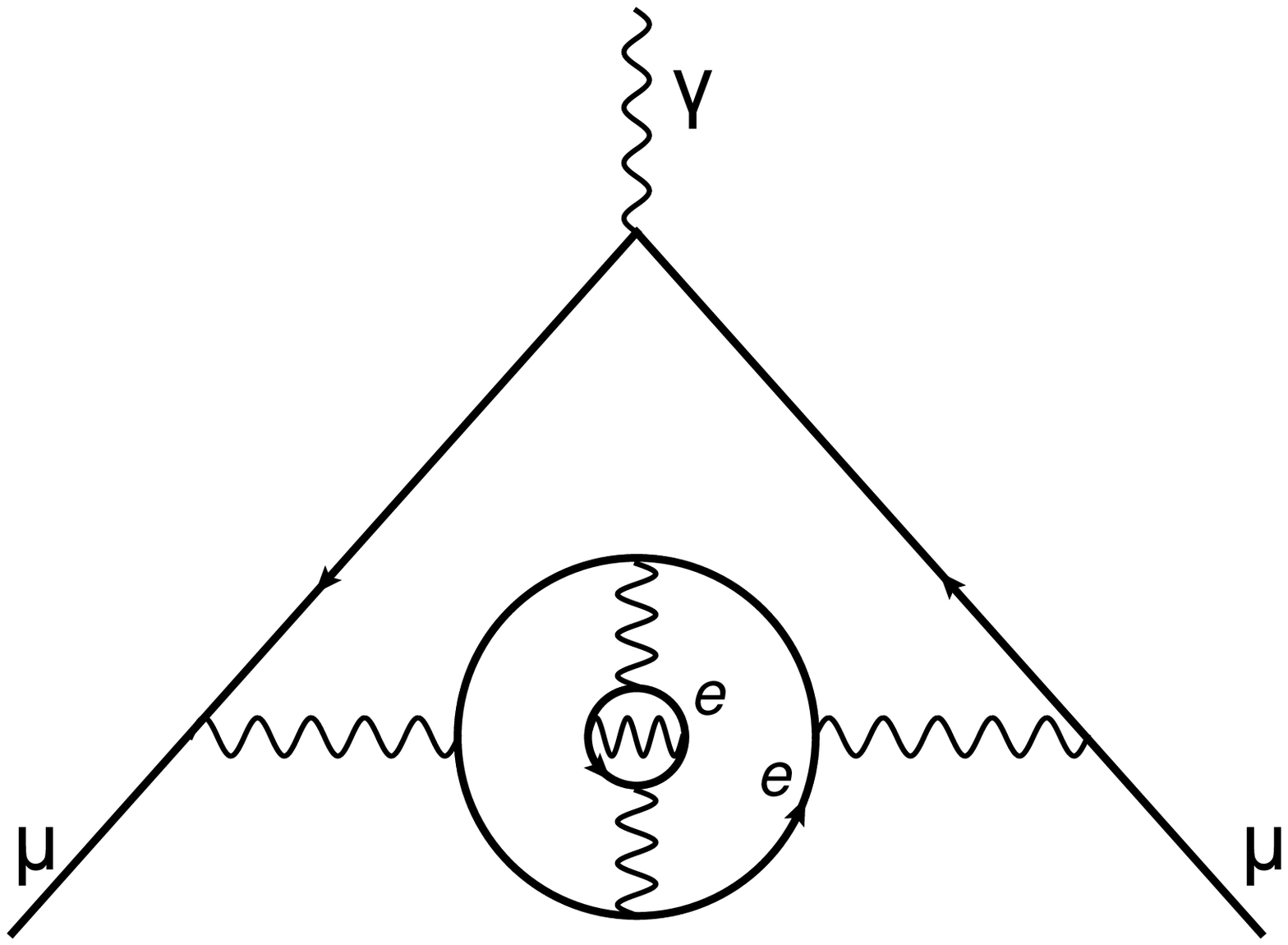}
\begin{center}
\vspace{-0.7cm}
{\tiny$I(g)$}
\end{center}
\end{minipage}
%
\begin{minipage}{3cm}
\includegraphics[bb=72 378 540 720,width=3cm]{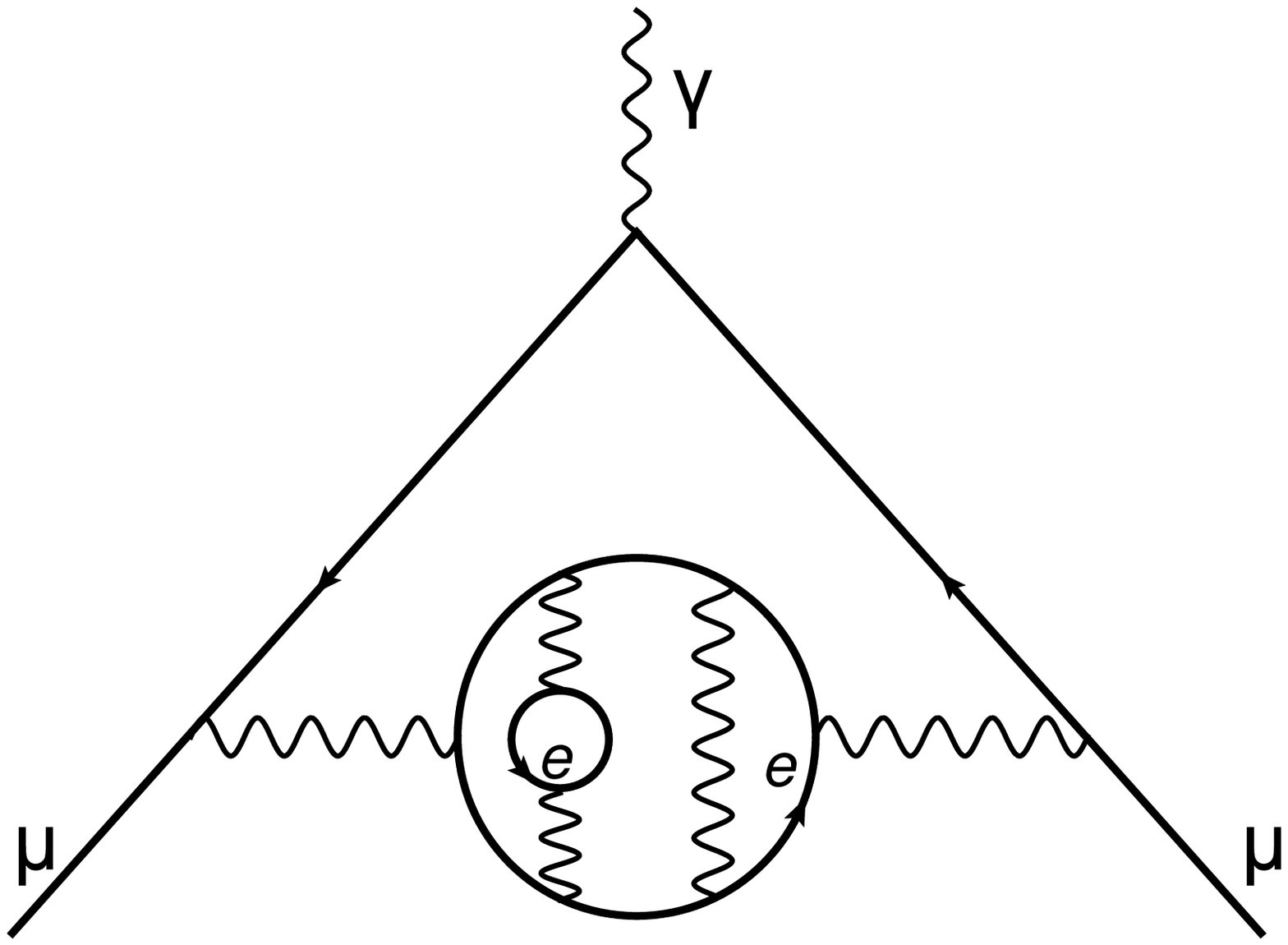}
\begin{center}
\vspace{-0.7cm}
{\tiny$I(h)$}
\end{center}
\end{minipage}
%
\begin{minipage}{3cm}
\includegraphics[bb=72 378 540 720,width=3cm]{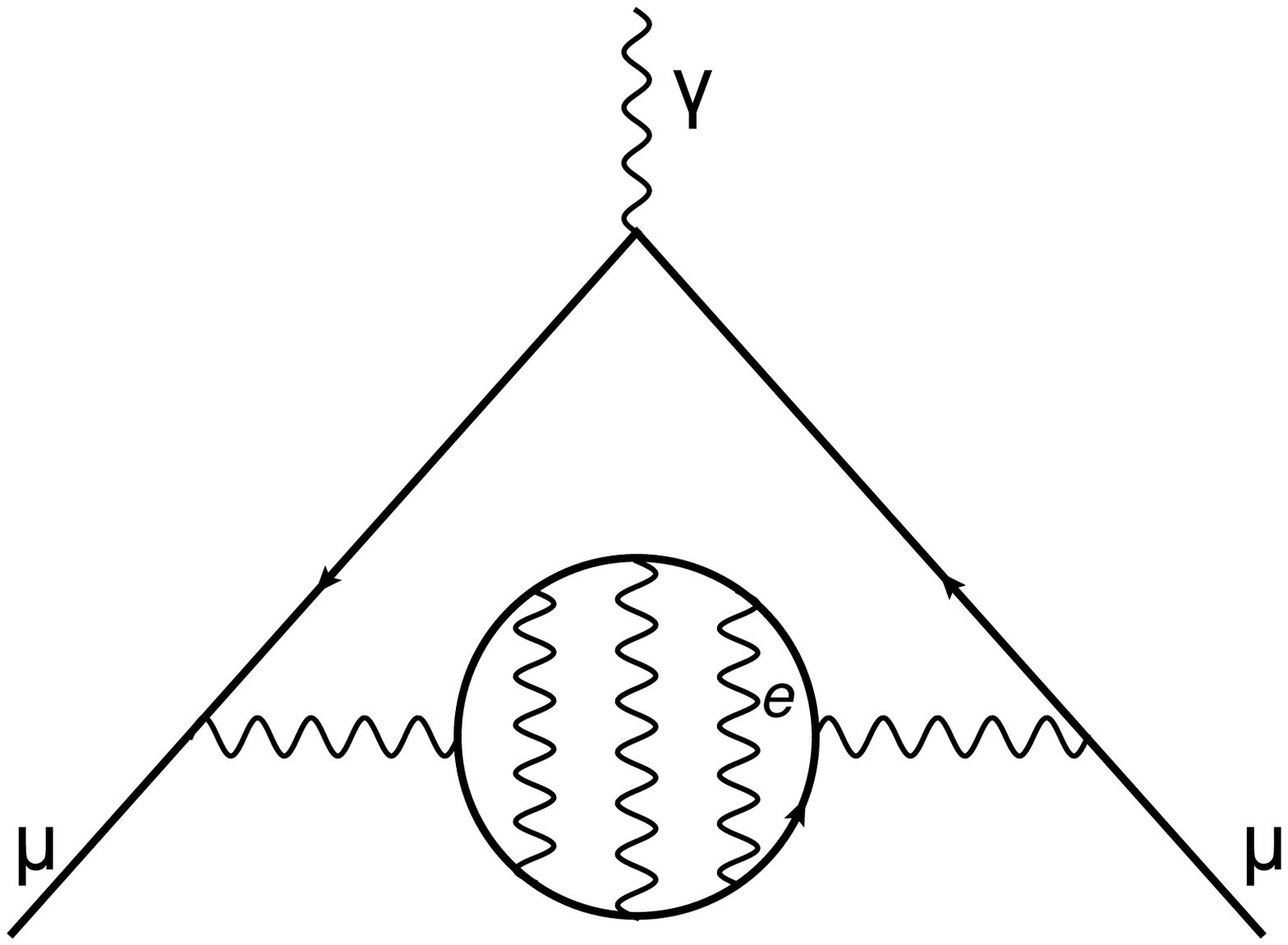}
\begin{center}
\vspace{-0.7cm}
{\tiny$I(i)$}
\end{center}
\end{minipage}
%
\begin{minipage}{3cm}
\includegraphics[bb=72 377 540 720,width=3cm]{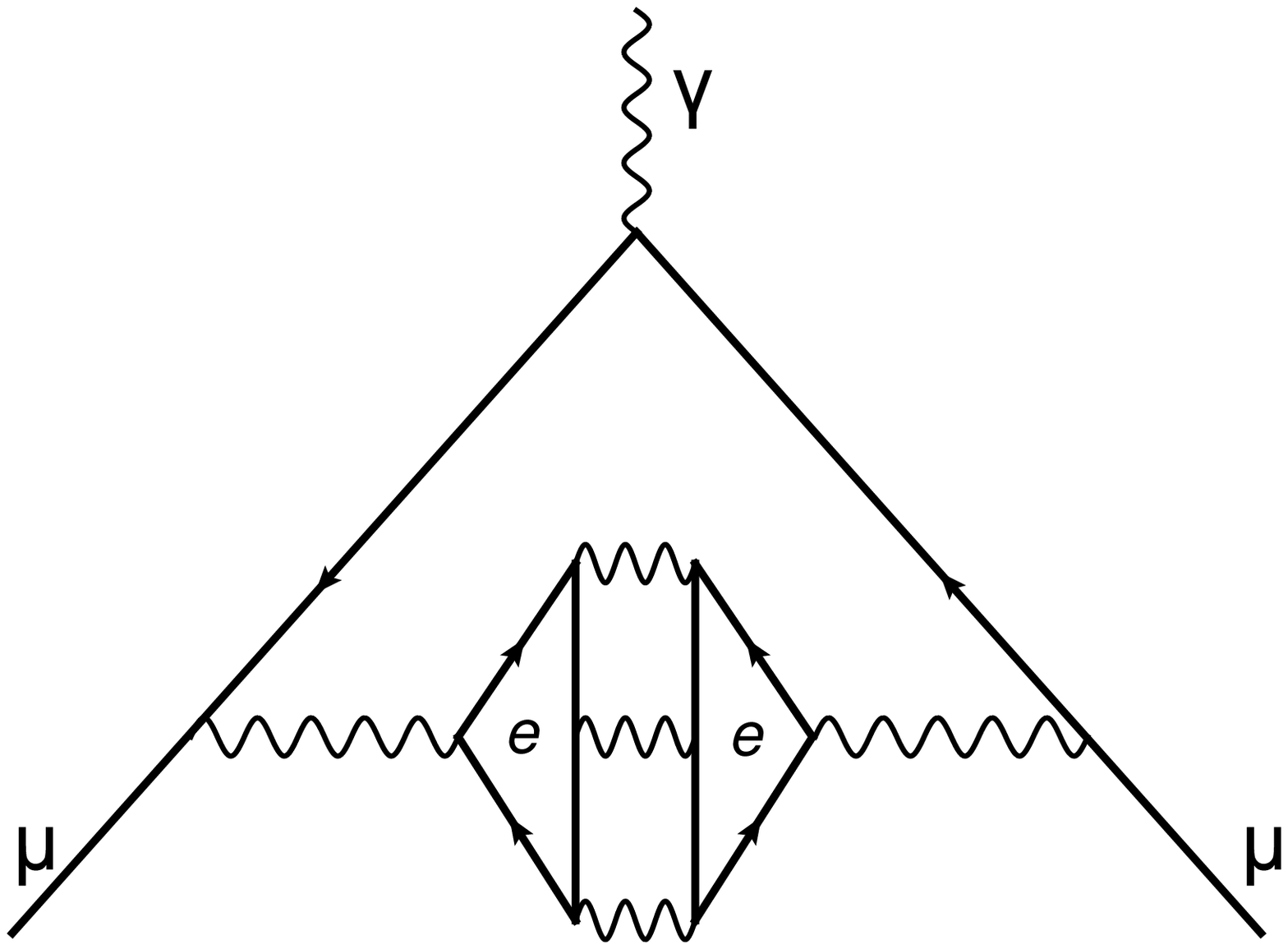}
\begin{center}
\vspace{-0.7cm}
{\tiny$I(j)$}
\end{center}
\end{minipage}
\end{center}
\vspace{-0.3cm}
\caption{\label{fig:MuonClasses} The ten gauge invariant subsets
  contributing to the muon anomaly which originate from inserting the
  vacuum polarization up to four-loop order into the first order QED
  vertex of Fig.~\ref{fig:classes}(a).  For each diagram class only one typical representative is
  shown.  Wavy lines denote photons($\gamma$), solid lines denote
  electrons($e$) or muons ($\mu$). The last five diagrams \{$I(f)$,
  $I(g)$, $I(h)$, $I(i)$, $I(j)$\} are non-factorizable insertions of the vacuum
  polarization function; the first five diagrams \{$I(a)$, $I(b)$,
  $I(c)$, $I(d)$, $I(e)$\} are factorizable ones.}
\end{figure}

Having at hand the result for the asymptotic polarization function at
four-loop order in QED, one can use it in combination with
Eqs.~(\ref{eq:dR}) and (\ref{eq:amuasymp}) to determine the full
asymptotic contribution (that is coming from electron vacuum polarization  insertions 
in  the  limit  $M_e/M_{\mu} \to 0$)
to the muon anomaly at
five-loops. The first of the six supersets of diagrams, which contribute
at five-loop order to the muon anomaly is shown in
Fig.~\ref{fig:MuonClasses}. It can be subdivided into ten gauge
invariant subsets.

\noindent
We define the perturbative expansion of the anomalous magnetic moment of
the muon by:
\begin{equation}
a^{\scriptsize asymp}_\mu= 
\sum_{i} a_{\mu}^{\scriptsize asymp,(i)}\left(\frac{\alpha}{\pi}\right)^i
{},\quad i=2,3,\dots.
\end{equation}
The decomposition of $a_\mu^{\scriptsize asymp}$ into the ten subsets shown
in Fig.~\ref{fig:MuonClasses} is given by:
\begin{eqnarray}
\label{eq:decompamu}
\left.a^{\scriptsize asymp,(5)}_\mu\right|_{Fig.~\ref{fig:MuonClasses}}&=&
 \amu{(5)}{a}+\amu{(5)}{b}+\amu{(5)}{c}+\amu{(5)}{d}+\amu{(5)}{e}
\nonumber\\
&+&\amu{(5)}{f}+\amu{(5)}{g+h}+\amu{(5)}{i}+\amu{(5)}{j}
{}.
\end{eqnarray}
The letters in the curly brackets denote the respective
    contribution of Fig.~\ref{fig:MuonClasses}.
Inserting the results of Eqs.~(\ref{eq:expPiOS})-(\ref{PiOs4}) into
Eq.~(\ref{eq:amuasymp}) and integrating
the different contributions of Eq.~(\ref{eq:decompamu}) we find:
\begin{eqnarray}
\label{eq:Ia}
\amu{(5)}{a}&=&
  {64613\over 26244 }
+ {317\over 729 }\*\pi^2
+ {2\over 135 }\*\pi^4
+ {100\over 81}\*\z3
+ \lmusdmes\*\left(-{8609\over 2187} - {100\over243}\*\pi^2 -
  {16\over27}\*\z3\right) 
\nonumber\\&&
+ \lmusdmes^2\*\left({634\over 243} + {8\over81}\*\pi^2\right) 
- \lmusdmes^3\*{200\over 243 }
+ \lmusdmes^4\*{8\over 81 }
,\\
\label{eq:Ib}
\amu{(5)}{b}&=&
- {439\over 162 }
- {35\over 108}\*\pi^2
+ {263\over 108}\*\z3
+ {\pi^2\over 9}\*\z3
+ \lmusdmes\*\left({413\over 108} + {\pi^2\over6} - {25\over 9}\*\z3\right)
\nonumber\\&&
+ \lmusdmes^2\*\left(-{35\over18} + {2\over3}\*\z3\right) 
+ {1\over 3 }\*\lmusdmes^3
,\\
\label{eq:Ic}
\amu{(5)}{c}&=&
  {409 \over 1152 }
+ {\pi^2\over 48 }
- {5\over 6 }\*\z3
+ {1\over 2}\*\z3^2
+ \lmusdmes\*\left(-{5\over 12} + {\z3\over 2}\right) 
+ {1\over 8}\*\lmusdmes^2 
,\\
\label{eq:Id}
\amu{(5)}{d}&=&
- {40057\over 15552 }
- {221\over 648 }\*\pi^2
+ {24185\over 10368 }\*\z3
+ {2\over 27}\*\pi^2\*\z3
+ \lmusdmes\*\left({3949\over1296} + {7\over 54}\*\pi^2 
                   - {1825\over864}\*\z3\right) 
\nonumber\\&&
+ \lmusdmes^2\*\left(-{121\over 108} + {4\over 9}\*\z3\right) 
+ {1\over 9}\*\lmusdmes^3
,\\
\label{eq:Ie}
\amu{(5)}{e}&=&
- {3409\over 3456 }
- {8\over 27}\*\pi^2
+ {25\over 54 }\*\pi^2\*\logtwo
- {275\over 128 }\*\z3
+ {125\over36 }\*\z5
\nonumber\\&&
+ \lmusdmes\*\left({161\over 288 }
                  + {5\over 36 }\*\pi^2
                  + {33 \over 32 }\*\z3
                  - {5 \over 3 }\*\z5
                  - {8\over 36}\*\pi^2\*\logtwo\right)
- {1 \over 24 }\*\lmusdmes^2
,\\
\label{eq:If}
\amu{(5)}{f}&=&
- {315079 \over 136080 }
- {34 \over 405 }\*\pi^2
+ {68869 \over 45360 }\*\z3
+ {\pi^2 \over 27 }\*\z3
+ {5 \over 18}\*\z5
\nonumber\\&&
+ \lmusdmes\*\left({152\over 81}+{\pi^2 \over 54}-{34\over27}\*\z3\right) 
+ \lmusdmes^2\*\left(-{4 \over 9} + {2\over9}\*\z3\right) 
+ {1\over 27}\*\lmusdmes^3
,\\
\label{eq:Igh}
\amu{(5)}{g+h}&=&
- {27413 \over 14400 }
- {53 \over 5 }\*\Pa4
- {53 \over 120 }\*\Log{2}{4}
+ {367 \over 1296 }\*\pi^2
- {8 \over 27 }\*\pi^2\*\logtwo
\nonumber\\&&
+ {53 \over 120 }\*\pi^2\*\Log{2}{2}
+ {2161 \over 21600 }\*\pi^4
- {18127 \over 1800 }\*\z3
+ {\z3^2\over 2 }
+ {50 \over 9}\*\z5
\nonumber\\&&
+ \lmusdmes\*\left(-{1\over 48} + {19\over12}\*\z3 - {5\over3}\*\z5\right) 
+ {1\over 24}\*\lmusdmes^2
,\\
\label{eq:Ii}
\amu{(5)}{i}&=&
  { 43357 \over 34560 }
- {1559 \over 90 }\*\Pa4
- {128 \over 15 }\*\Pa5
- {1559 \over 2160}\*\Log{2}{4}
+ {16 \over 225 }\*\Log{2}{5}
+ {157 \over 144 }\*\pi^2
\nonumber\\&&
- {59 \over 24 }\*\pi^2\*\logtwo
+ {1559 \over 2160 }\*\pi^2\*\Log{2}{2}
- {16 \over 135 }\*\pi^2\*\Log{2}{3}
+ {59801 \over 259200 }\*\pi^4
\nonumber\\&&
- {53 \over 675 }\*\pi^4\*\logtwo
- {6559 \over 640 }\*\z3
+ {\pi^2 \over 48 }\*\z3
+ {1603 \over 240 }\*\z5
+ {35 \over 8}\*\z7
- {23 \over 128 }\*\lmusdmes
,\\
\label{eq:Ij}
\amu{(5)}{j}&=&
- {9701\over 15120 }
- {73 \over 12 }\*\Pa4
- {73 \over 288 }\*\Log{2}{4}
+ {73 \over 288 }\*\pi^2\*\Log{2}{2}
+ {2237 \over 34560 }\*\pi^4
\nonumber\\&&
- {10327 \over 6720 }\*\z3
+ {\z3^2 \over 3} 
- {5 \over 6}\*\z5
+ \lmusdmes\*\left({11 \over 36} - {2 \over 3}\*\z3 \right) 
,
\end{eqnarray}
with $\lmusdmes=\ln({M_\mu/M_e})$. Power suppressed terms of the
order $\mathcal{O}(M_e/M_\mu)$ are neglected.
The coefficients of the logarithmic terms $\lmusdmes$ have been
computed analytically in Ref.~\cite{Kataev:1991cp}, except for the
complete $\lmusdmes$-term in Eq.~(\ref{eq:Ie}). For the factorizable
insertions of the vacuum polarization function also the mass independent
term has been determined analytically in
Refs.~\cite{Kataev:1991cp,Laporta:1994md}, except for the case
$I(e)$. All the analytical results of
Refs.~\cite{Kataev:1991cp,Broadhurst:1992za,Laporta:1994md} for 
Eqs.~(\ref{eq:If})-(\ref{eq:Ia}) are in agreement with ours. Numerical
results for these ten gauge invariant subsets of diagrams which are
shown in Fig.~\ref{fig:MuonClasses} have been reported in
Refs.~\cite{Kinoshita:2005sm,Aoyama:2008hz,Aoyama:2010zp,Aoyama:2008gy}.
A comparison between the asymptotic analytical formulas of
Eqs.~(\ref{eq:Ia})-(\ref{eq:If}) and the numerical results is shown in
Table~\ref{tab:comparison}.
\begin{table}[!ht]
\begin{center}
\begin{tabular}{|c||rl|ll|l|}
\hline
{Subset}& \multicolumn{2}{|c|}{Analytical} &
\multicolumn{1}{|c}{Numerical}& Ref.& Num.-ana.\\
\hline
$I(a)$         & 20.1832 &\hspace{-0.25cm}+ $\mathcal{O}(M_e/M_\mu)$ &\ph 20.14293(23) & \cite{Kinoshita:2005sm}&$\approx$ -0.04\\
$I(b)$         & 27.7188 &\hspace{-0.25cm}+ $\mathcal{O}(M_e/M_\mu)$ &\ph 27.69038(30) & \cite{Kinoshita:2005sm}&$\approx$ -0.03\\
$I(c)$         & 4.81759 &\hspace{-0.25cm}+ $\mathcal{O}(M_e/M_\mu)$ &\ph 4.74212(14) & \cite{Kinoshita:2005sm}&$\approx$ -0.08\\
$I(d)$         & 7.44918 &\hspace{-0.25cm}+ $\mathcal{O}(M_e/M_\mu)$ &\ph 7.45173(101) & \cite{Kinoshita:2005sm}&$\approx$\hspace*{0.7ex}  0.003\\
$I(e)$         &-1.33141 &\hspace{-0.25cm}+ $\mathcal{O}(M_e/M_\mu)$ & -1.20841(70) &\cite{Kinoshita:2005sm}&$\approx$ \ph 0.12\\\hline\hline
$I(f)$         & 2.89019 &\hspace{-0.25cm}+ $\mathcal{O}(M_e/M_\mu)$ &\ph 2.88598(9) & \cite{Kinoshita:2005sm}&$\approx$ -0.004\\
$I(g) + I(h)$  & 1.50112 &\hspace{-0.25cm}+ $\mathcal{O}(M_e/M_\mu)$ &\ph 1.56070(64)& \cite{Aoyama:2008hz}&$\approx$ \ph 0.06\\
$I(i)$         & 0.25237 &\hspace{-0.25cm}+ $\mathcal{O}(M_e/M_\mu)$ &\ph 0.0871(59) & \cite{Aoyama:2010zp}&$\approx$ -0.17\\
$I(j)$         &-1.21429 &\hspace{-0.25cm}+ $\mathcal{O}(M_e/M_\mu)$ &   -1.24726(12)& \cite{Aoyama:2008gy}&$\approx$ -0.03\\
\hline
\end{tabular}
\end{center}
\vspace{-0.6cm}
\caption{\label{tab:comparison}The first column shows the different
  gauge invariant subsets of diagrams as defined in
  Fig.~\ref{fig:MuonClasses}. The second column contains the
  corresponding results of Eqs.~(\ref{eq:Ia})-(\ref{eq:Ij}) evaluated
  numerically, where we have used for the mass ratio
  $M_\mu/M_e=206.7682843(52)$\cite{Mohr:2012tt}. This result is correct
  only up to power corrections in the small mass ratio $M_e/M_\mu$. The
  third column contains the numerical result obtained in
  Refs.~\cite{Kinoshita:2005sm,Aoyama:2008hz,Aoyama:2010zp,Aoyama:2008gy}.
  (Note that  in Ref.~\cite{Kinoshita:2005sm} also a more precise value $7.45270(88)$
 is given for $I(d)$ which was obtained using the exact sixth order spectral function.)
 The last column shows the difference between the numerical and
  asymptotic analytical results. The subsets \{$I(a)$, $I(b)$,
  $I(c)$, $I(d)$, $I(e)$\} originate from Feynman
  diagrams with factorizable vacuum polarization insertions, whereas the
  subsets \{$I(f)$, $I(g)$, $I(h)$, $I(i)$, $I(j)$\} are non-factorizable
  (see Fig.~\ref{fig:MuonClasses}).} 
\end{table}

\noindent
In general good agreement between asymptotic analytical and numerical
results is found, except for the subsets $I(e)$ and $I(i)$ where we only
have poor agreement. The remaining differences should arise from
corrections of the order $\mathcal{O}(M_e/M_\mu)$.%
\footnote{For a discussion of the differences for subset $I(i)$ see also Section VIII of Ref.~\cite{Aoyama:2010zp}.}
Summing up all ten subsets one obtains $\left.a^{\scriptsize
  asymp,(5)}_\mu\right|_{Fig.~\ref{fig:MuonClasses}}=62.26675+\mathcal{O}(M_e/M_\mu)$
which is, despite  the small number of diagrams, sizeable
($\approx8$\%) compared to the complete numerical result
$a_\mu^{(5)}=753.29(1.04)$ of Ref.~\cite{Aoyama:2012wk}. The reason for
this is the logarithmic enhancement.
Indeed, if we set to zero all $Q$-dependent terms in the asymptotic
OS polarization function the result for 
$\left.a^{\scriptsize
  asymp,(5)}_\mu\right|_{Fig.~\ref{fig:MuonClasses}}
$ would be reduced to  2.52261.

\section{Contributions to the muon anomaly at six loops\label{sec:muon6}}

The  five-loop contribution to Eq.~(\ref{eq:dR})
\[
d^{\mbox{\scriptsize asymp}}_R(Q^2/M^2,\alpha) = \sum_{i \ge 0} d_i(\ellMQ)\, \left(\frac{\al}{\pi}\right)^i
\]
can be schematically  represented as follows:
\beq
\label{eq:d5decomp}
d_5 = d_5^{\mathrm{factr}} - \Pi^{\asy , (5)} + C_5
{},
\eeq
where 
\bea
d_5^{\mathrm{factr}} &=& -\left(\Pi^{\asy , (1)}\right)^5 
+ 4\, \left(\Pi^{\asy , (1)}\right)^3\,\Pi^{\asy , (2)} 
- 3\, \Pi^{\asy , (1)}\,\left(\Pi^{\asy , (2)}\right)^2
\nonumber
\\
&{-}&
  3\, \left(\Pi^{\asy , (1)}\right)^2 \,  \Pi^{\asy , (3)}
+ 2\, \Pi^{\asy , (2)} \,\Pi^{\asy , (3)} 
+ 2\,  \Pi^{\asy , (1)}\,  \Pi^{\asy , (4)}
{}
\label{eq:d5factr}
\eea
and $C_5$ is a still unknown constant standing for Q-independent contributions
missing in   $\Pi^{\asy , (5)}$ as described by Eq.~\re{PiOS5}.

The corresponding result for the asymptotic contribution to the muon anomaly 
is rather bulky so we provide it  below in numerical form only: 
\beq
a^{\scriptsize  asymp,(6)}_\mu = 257.245 + \frac{C_5}{2} =  246.381^{\mathrm{factr}} + 10.8647 + \frac{C_5}{2}
{},
\label{amu6}
\eeq
where after  the second equality sign  we  decompose the  result into two pieces:
the factorizable and the genuine six-loop terms.  
We observe that the six-loop asymptotic contribution to the 
muon anomaly is almost completely saturated by factorizable terms. One could
hardly expect that the constant $C_5$ might have any  numerical  relevance.
Indeed,  an analog of Eq.~\re{amu6} at the five-loop  level looks as:
\beq
a^{\scriptsize  asymp,(5)}_\mu = 60.7524+ \frac{C_4=3.0287}{2} 
=  58.8374^{\mathrm{factr}}  + 1.915     + \frac{C_4=3.0287}{2}   {}.
\label{amu5}
\eeq
Thus, we believe that Eq.~\re{amu6} with $C_5 = 0$ presents quite a good
prediction for   $a^{\scriptsize  asymp,(6)}_{\mu}$.
\begin{table}[!ht]
\begin{center}
\begin{tabular}{|c|c|c|c|c|c|c|}
\hline
{Subset}&$N^4$   &$N^3$   &$N^2$    &$N$      &${\bf si}\,N^3$ &${\bf si}\,N^2$\\\hline
{Value} & 7.15995& 7.15320& -2.73813& 3.37488&-7.22943        & 3.14426\\\hline
\end{tabular}
\end{center}
\caption{\label{tab:sixloopnonfact} The numerical values are correct up
  to power corrections of the order $\mathcal{O}\left(M_e/M_\mu\right)$
  and the contribution coming from the unknown constant $C_5$ of
  Eq.~(\ref{eq:d5decomp}). The symbol ${\bf si}$ labels singlet
  contributions, whereas the power of $N$ denotes the number of closed
  fermion loops which arise in the corresponding diagrams.}
\end{table}

\begin{figure}[!ht]
\begin{center}
\begin{minipage}{3.5cm}
\includegraphics[bb=72 437 540 720,width=3.5cm]{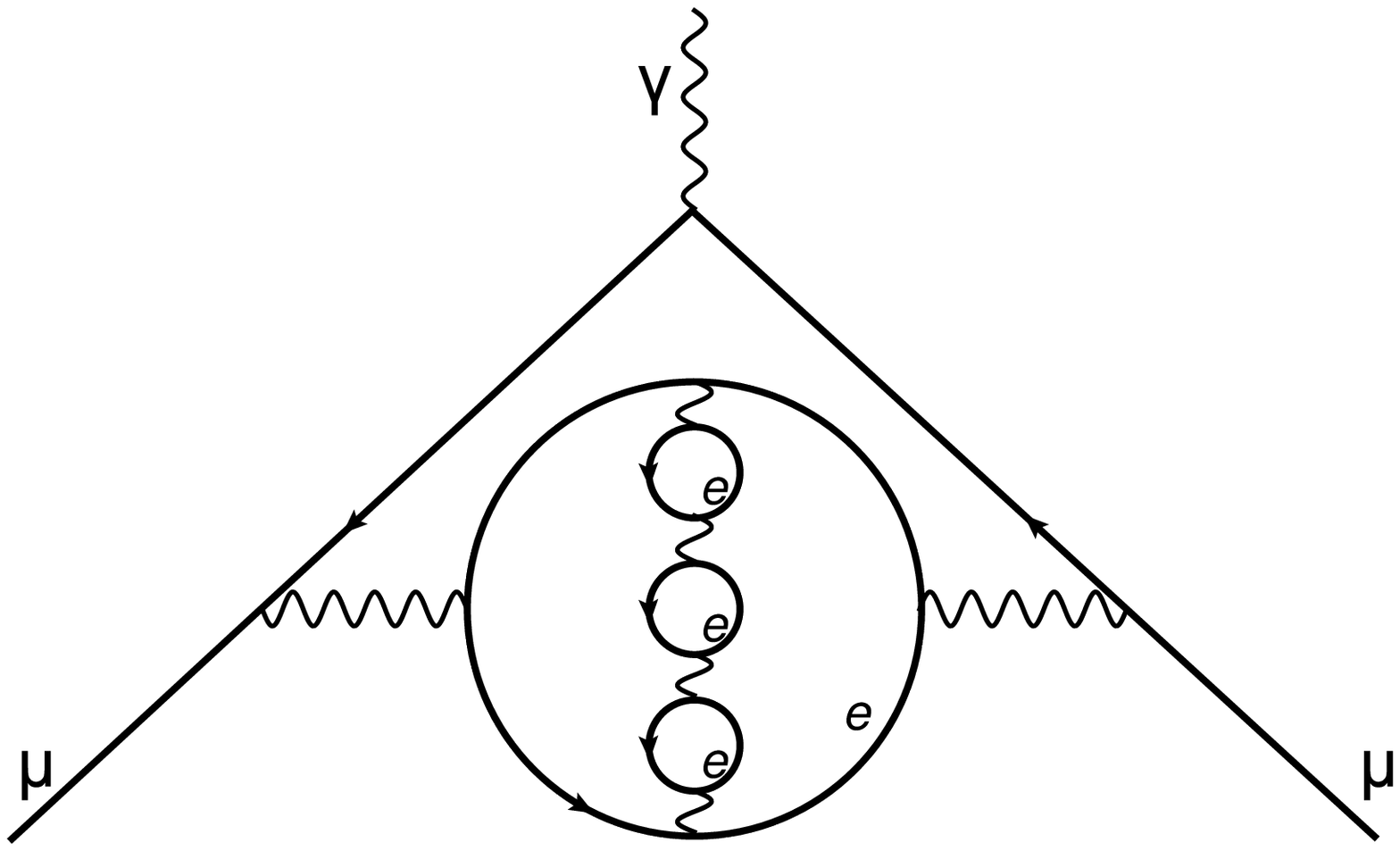}
\begin{center}
\vspace{-0.6cm}
{\tiny$N^4$}
\end{center}
\end{minipage}
\hspace{0.3cm}
\begin{minipage}{3.5cm}
\includegraphics[bb=72 437 540 720,width=3.5cm]{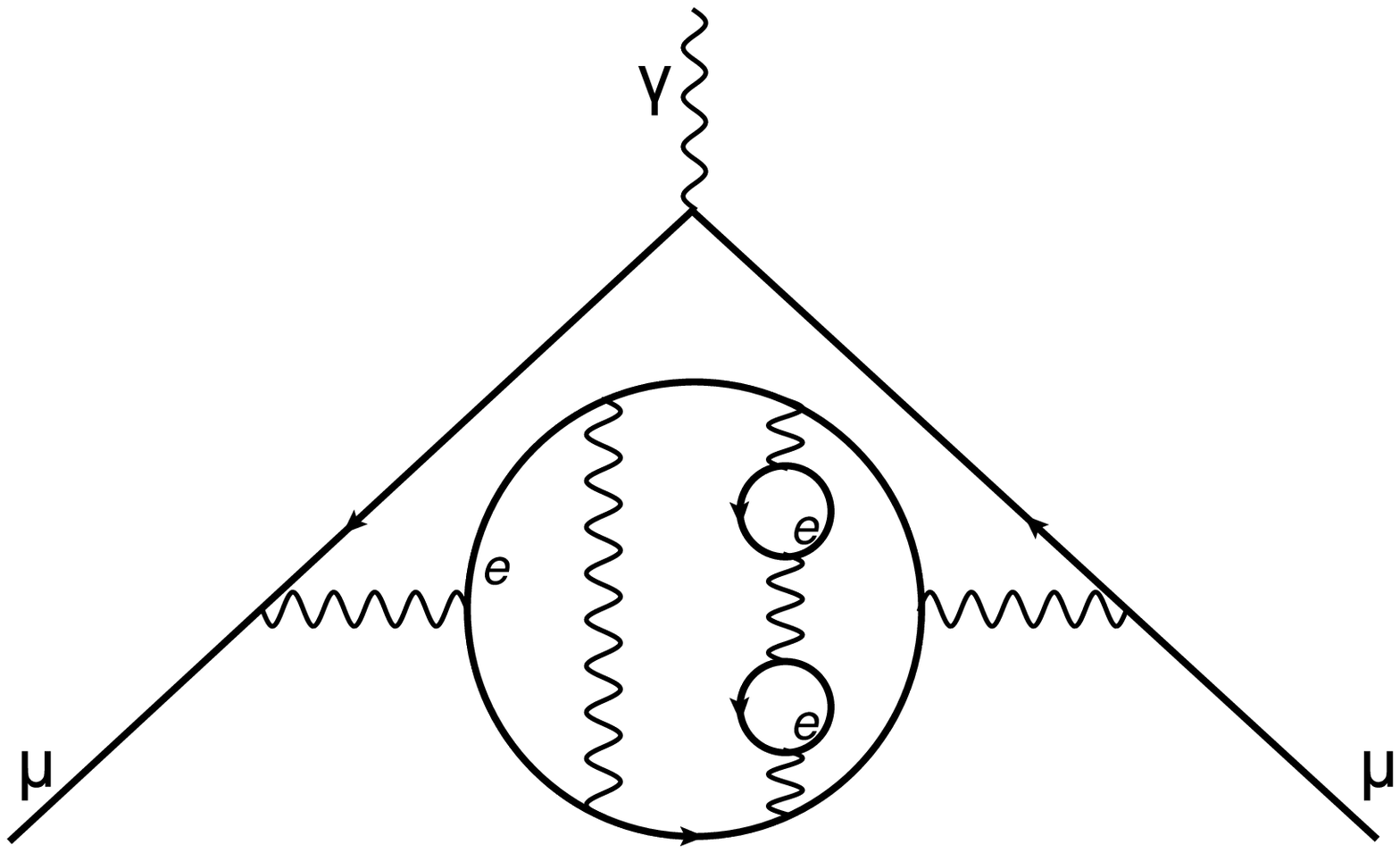}
\begin{center}
\vspace{-0.6cm}
{\tiny$N^3$}
\end{center}
\end{minipage}
\hspace{0.3cm}
\begin{minipage}{3.5cm}
\includegraphics[bb=72 437 540 720,width=3.5cm]{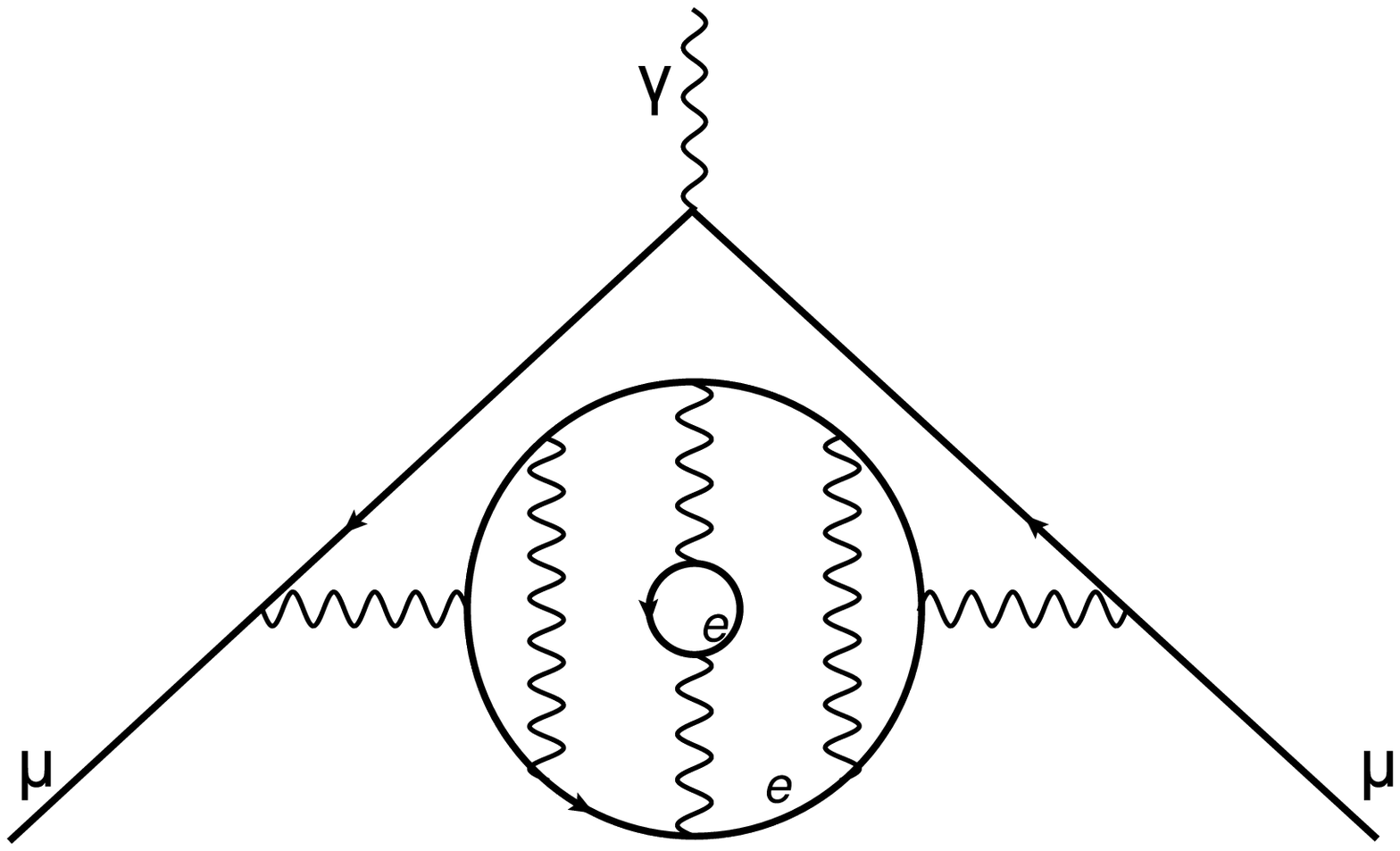}
\begin{center}
\vspace{-0.6cm}
{\tiny$N^2$}
\end{center}
\end{minipage}\\[0.3cm]
%
%
\begin{minipage}{3.5cm}
\includegraphics[bb=72 437 540 720,width=3.5cm]{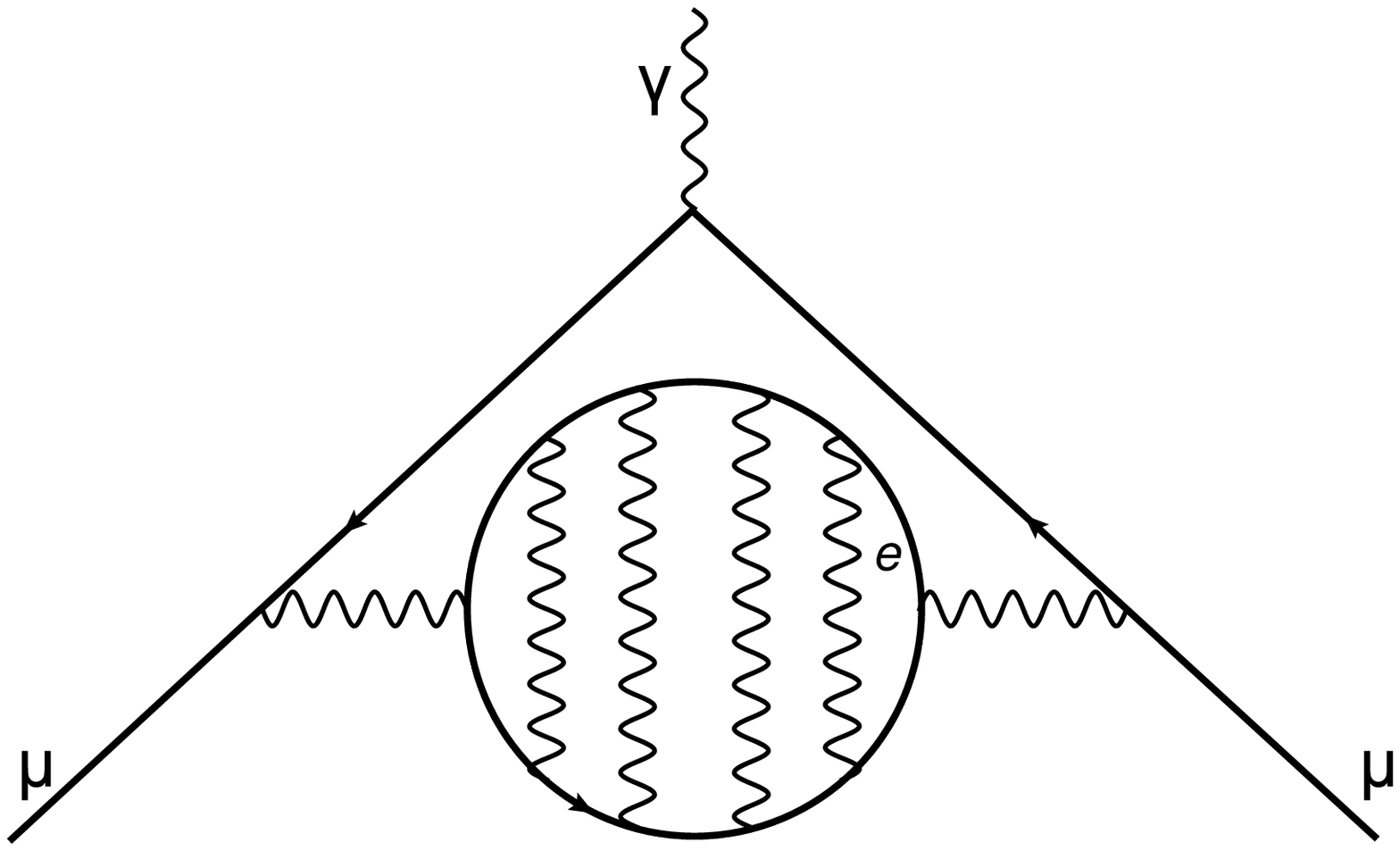}
\begin{center}
\vspace{-0.6cm}
{\tiny$N$}
\end{center}
\end{minipage}
\hspace{0.3cm}
\begin{minipage}{3.5cm}
\includegraphics[bb=72 437 540 720,width=3.5cm]{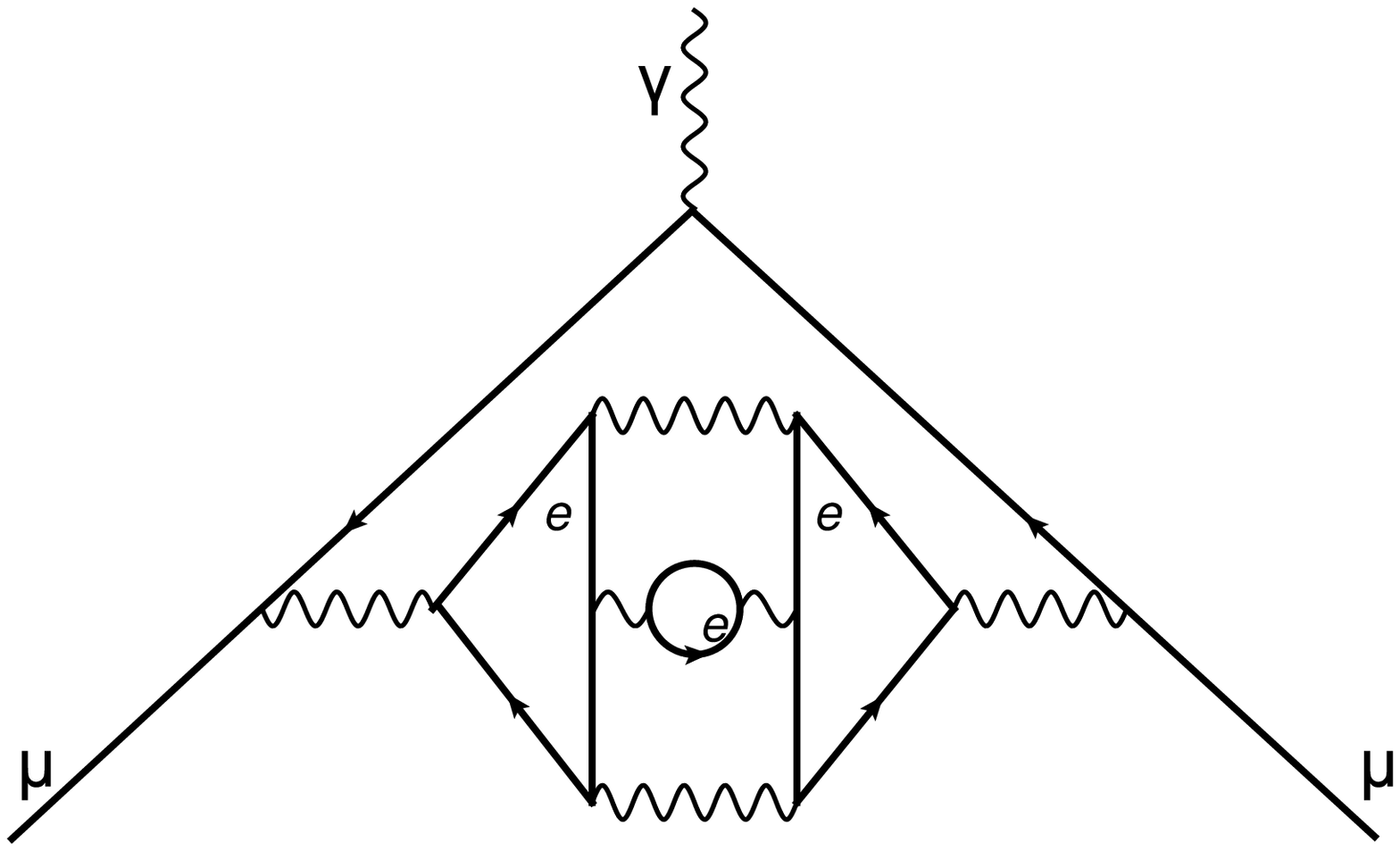}
\begin{center}
\vspace{-0.6cm}
{\tiny${\bf{si}}N^3$}
\end{center}
\end{minipage}
\hspace{0.3cm}
\begin{minipage}{3.5cm}
\includegraphics[bb=72 437 540 720,width=3.5cm]{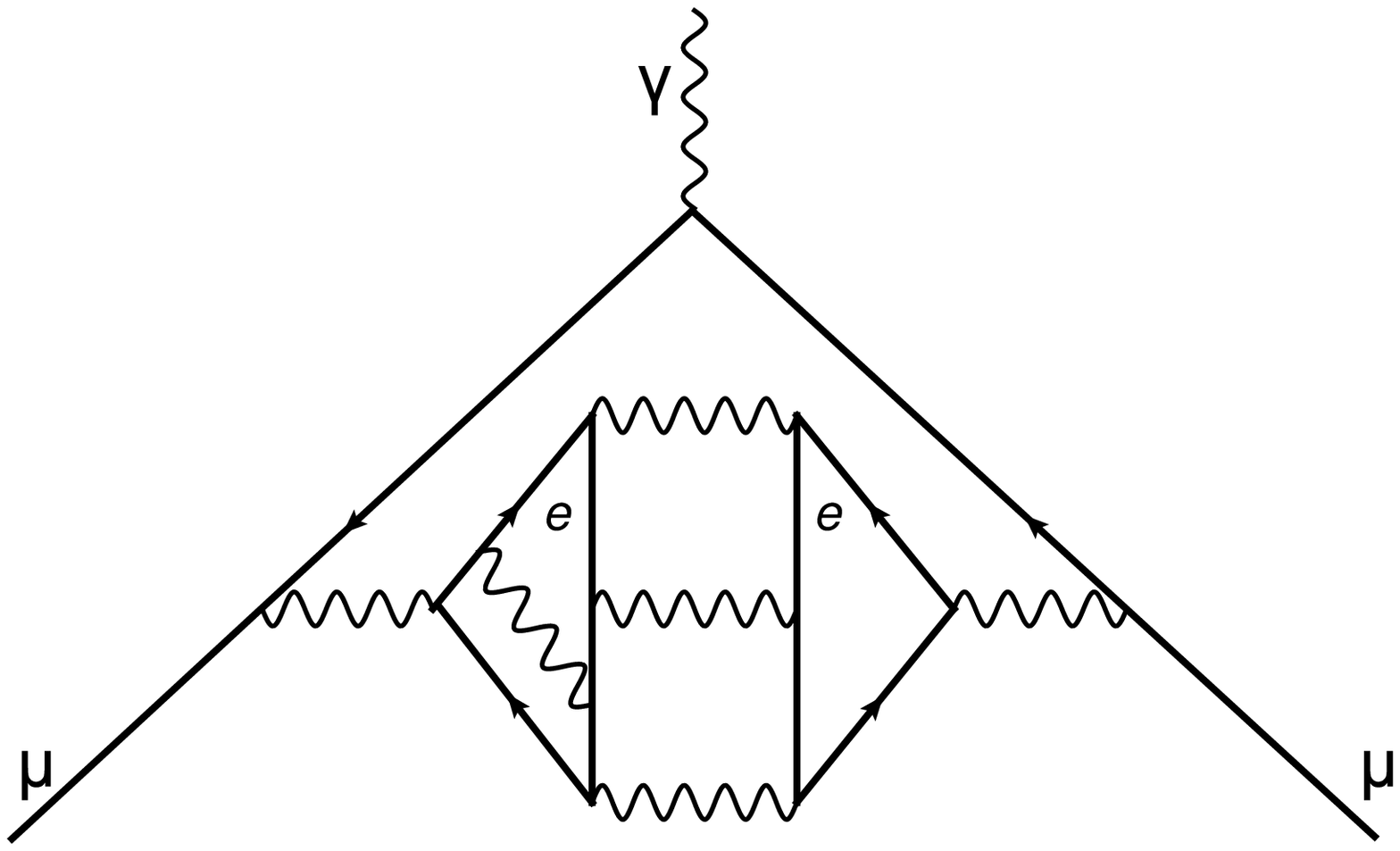}
\begin{center}
\vspace{-0.6cm}
{\tiny${\bf{si}}N^2$}
\end{center}
\end{minipage}
\end{center}
\vspace{-0.3cm}
\caption{\label{fig:MuonClasses6loop} For each subset of
  Table~\ref{tab:sixloopnonfact} one example diagram is shown.  It would
  be possible to subdivide these diagram classes into further subsets,
  however, we refrain from doing this for simplicity.  The symbol ${\bf
    si}$ denotes a singlet diagram, whereas the power of $N$ gives the
  number of closed fermion loops which arise in the corresponding
  diagrams. Wavy lines represent again photons whereas solid lines stand
  for electrons or muons.} 
\end{figure}

To further study the size of the factorizable and
non-factorizable six-loop contributions to the muon anomaly we
subdivide them into several gauge invariant subsets in complete analogy to
the five-loop case shown in Fig.~\ref{fig:MuonClasses}. Let us start
with the genuine, non-factorizable contributions which correspond to the
term $- \Pi^{\asy , 5} + C_5$ of Eq.~(\ref{eq:d5decomp}) and which we
again subdivide according to the number of fermion loops arising in the
corresponding diagrams. The individual contributions to
$\left.a^{\scriptsize asymp,(6)}_\mu\right|_{\mathrm{non-factr}}
=10.8647 + C_5/2$ of Eq.~(\ref{amu6}) are shown in numerical evaluated
form in Table~\ref{tab:sixloopnonfact} and an example diagram for each of
the six subsets of Table~\ref{tab:sixloopnonfact} is given in
Fig.~\ref{fig:MuonClasses6loop}.\\

\begin{figure}[!ht]
\begin{center}
\begin{minipage}{3.5cm}
\includegraphics[bb=72 436 540 720,width=3.5cm]{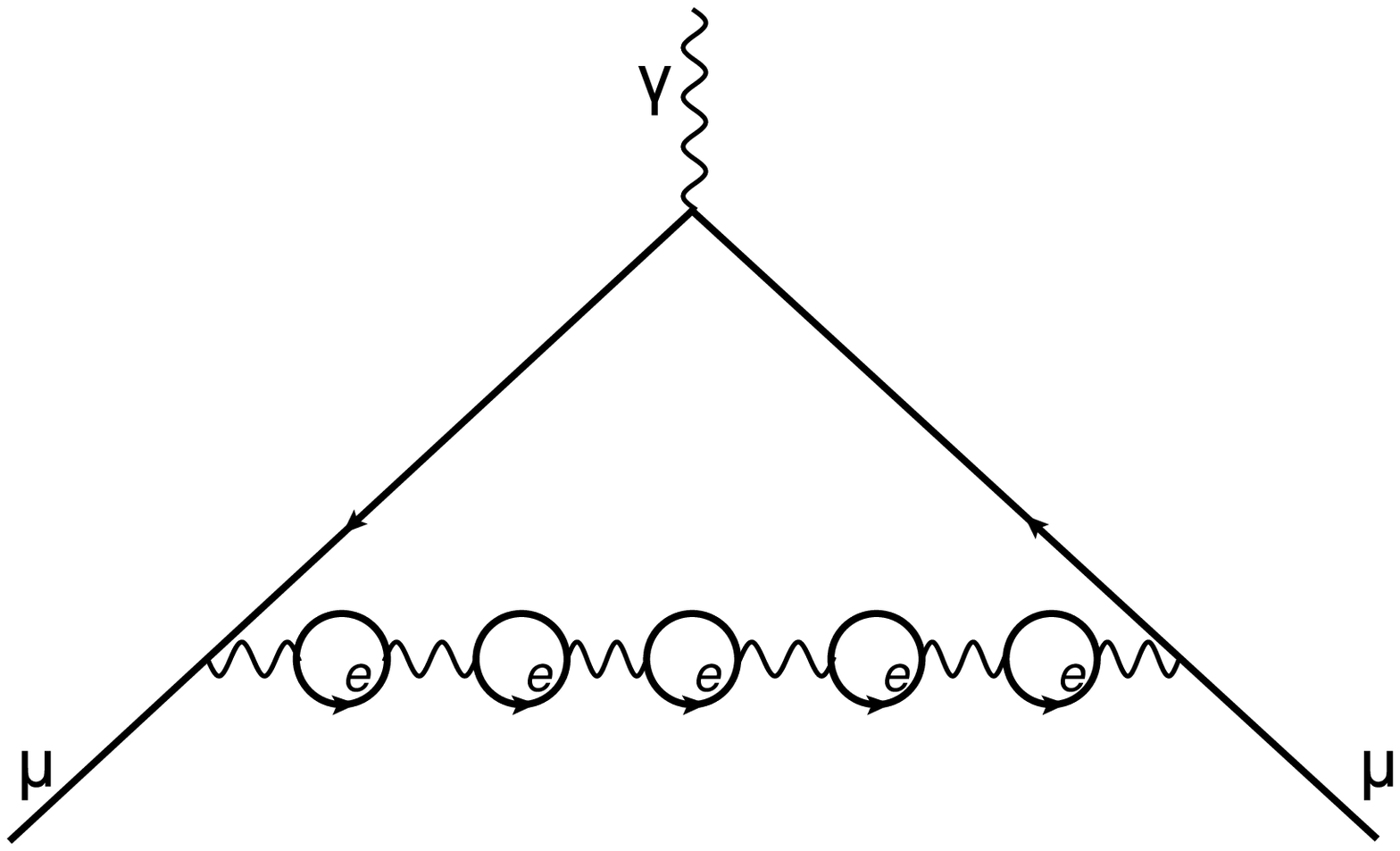}
\begin{center}
\vspace{-0.6cm}
{\tiny$I_6(a)$}
\end{center}
\end{minipage}
\hspace{0.2cm}
\begin{minipage}{3.5cm}
\includegraphics[bb=72 436 540 720,width=3.5cm]{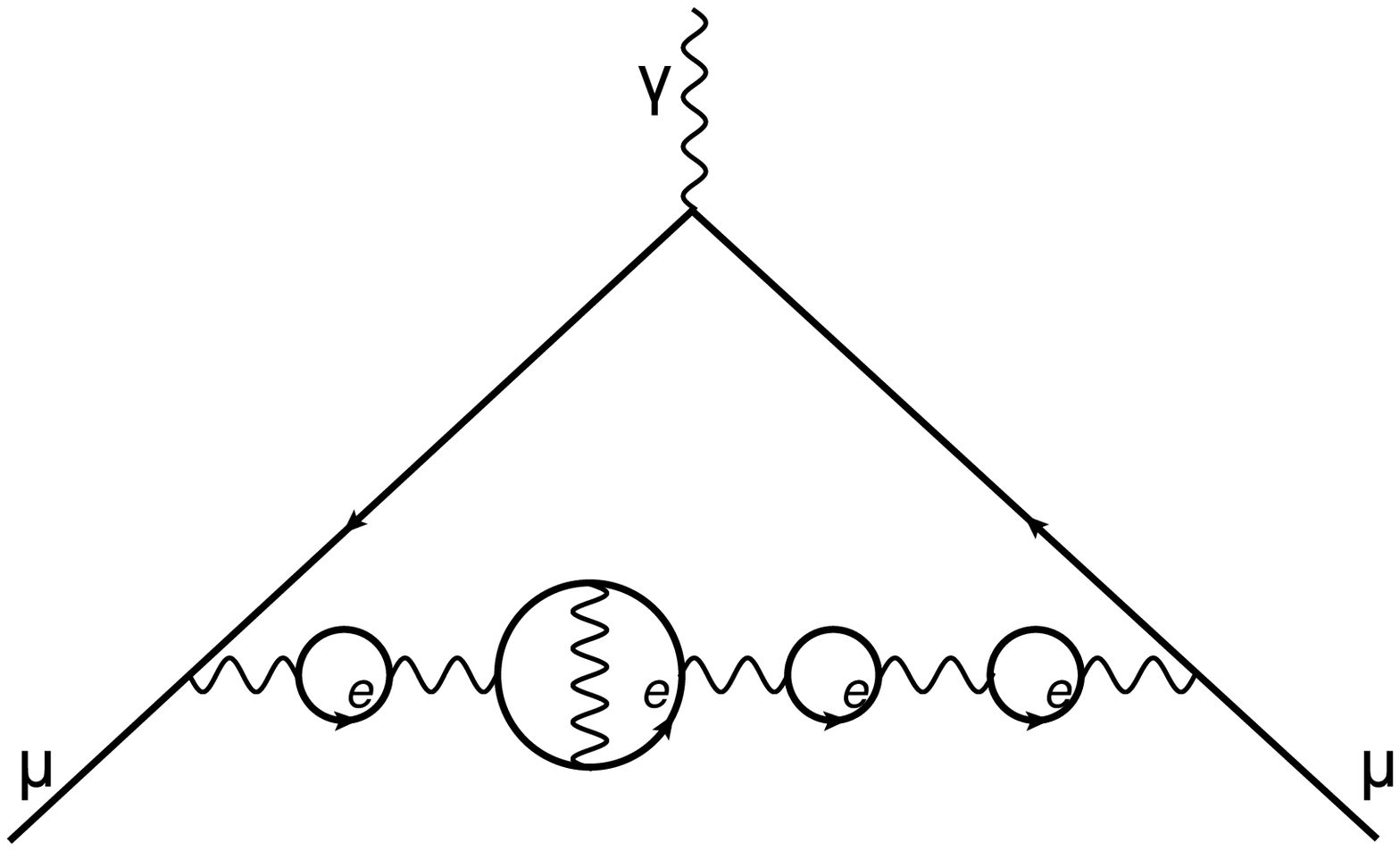}
\begin{center}
\vspace{-0.6cm}
{\tiny$I_6(b)$}
\end{center}
\end{minipage}
\hspace{0.2cm}
\begin{minipage}{3.5cm}
\includegraphics[bb=72 436 540 720,width=3.5cm]{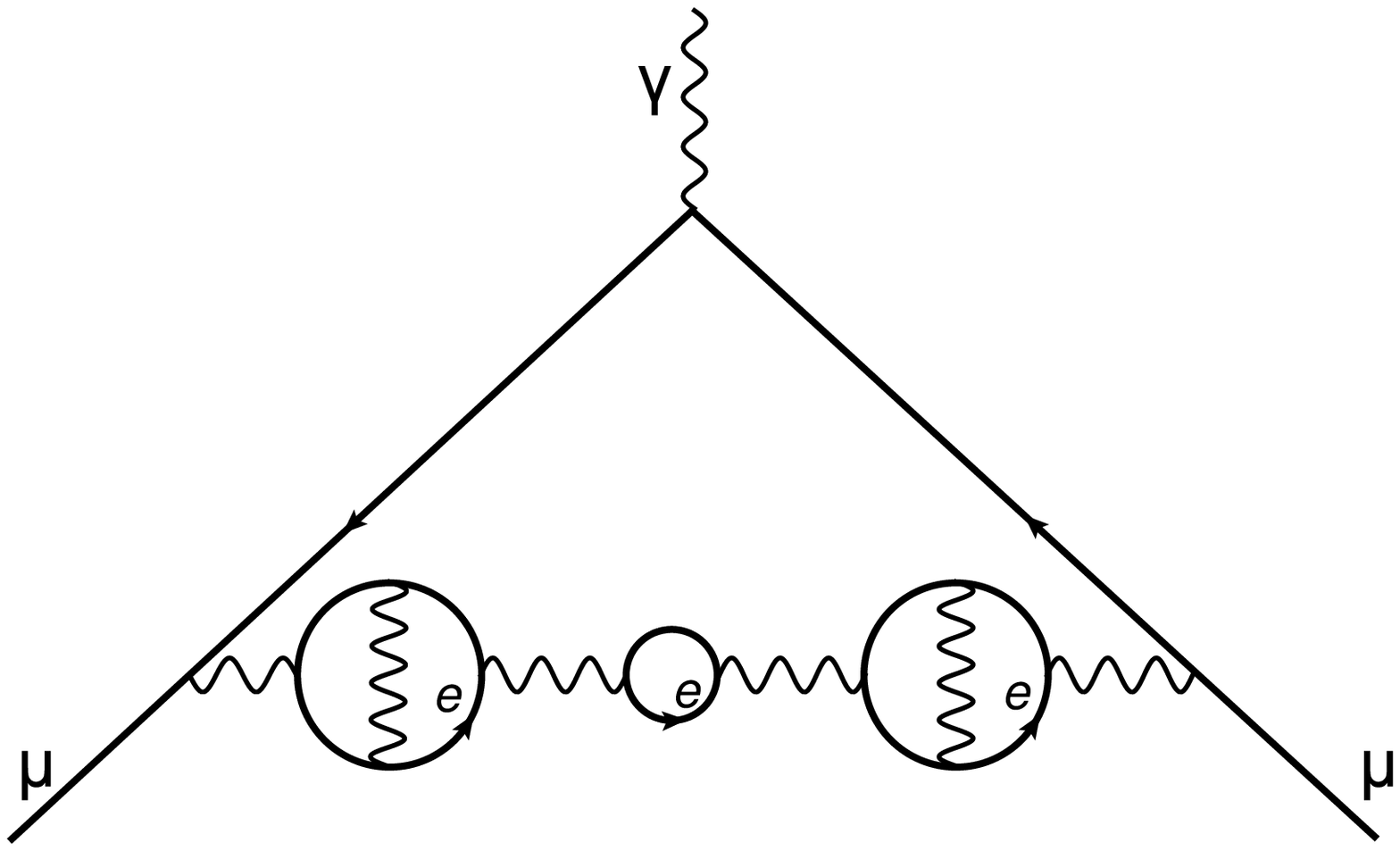}
\begin{center}
\vspace{-0.6cm}
{\tiny$I_6(c)$}
\end{center}
\end{minipage}
\hspace{0.2cm}
\begin{minipage}{3.5cm}
\includegraphics[bb=72 436 540 720,width=3.5cm]{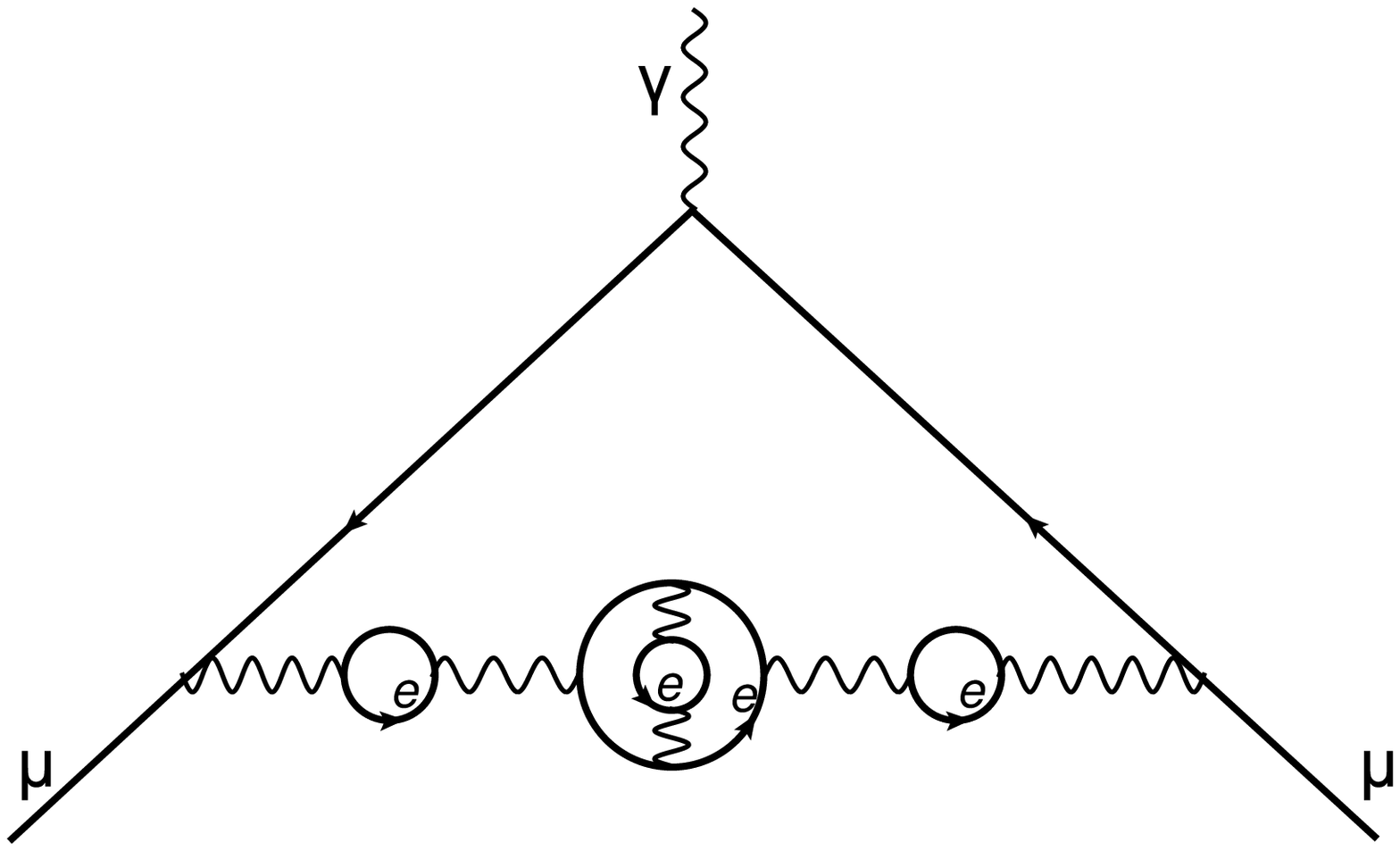}
\begin{center}
\vspace{-0.6cm}
{\tiny$I_6(d)$}
\end{center}
\end{minipage}\\[0.3cm]
\begin{minipage}{3.5cm}
\includegraphics[bb=72 436 540 720,width=3.5cm]{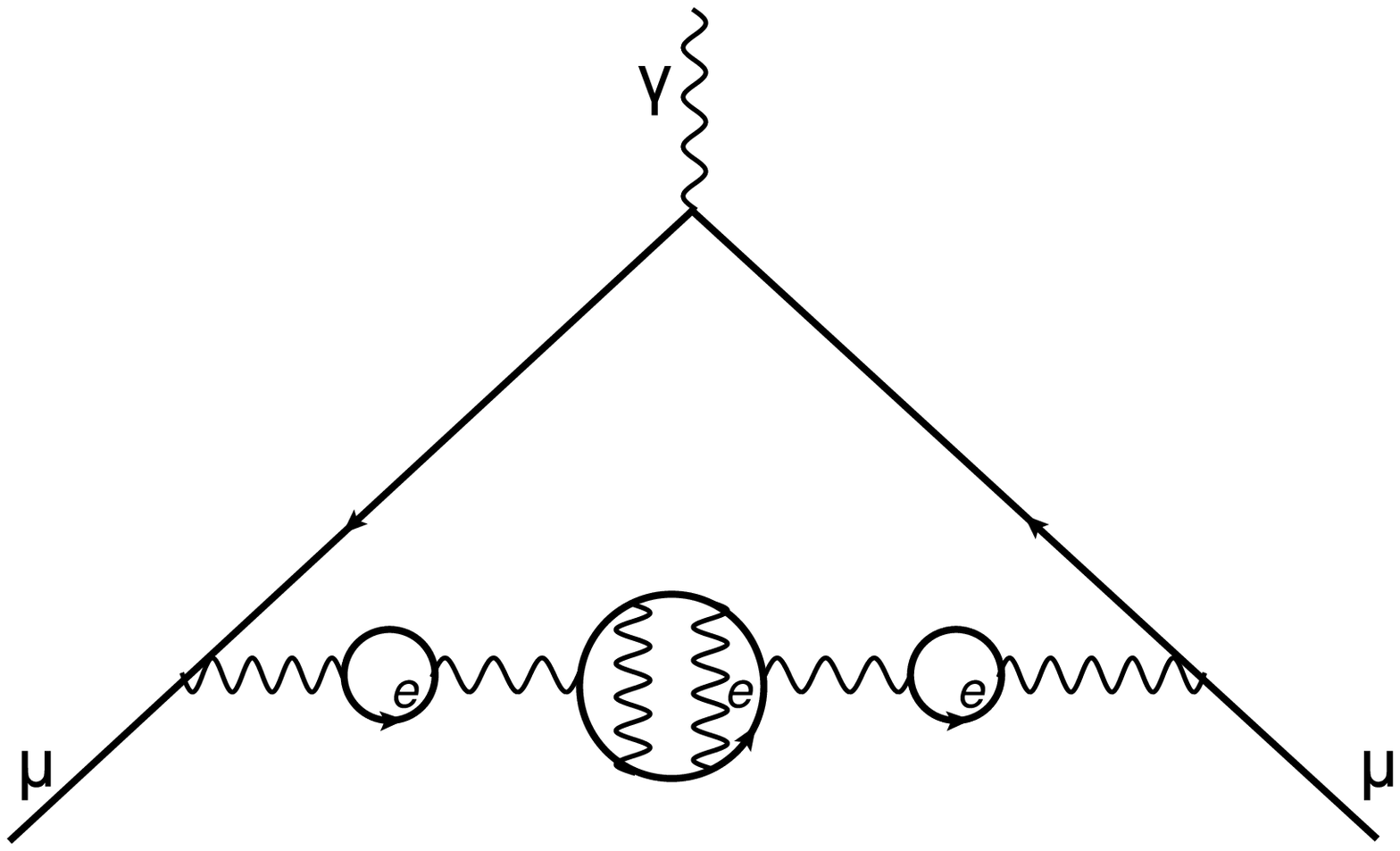}
\begin{center}
\vspace{-0.6cm}
{\tiny$I_6(e)$}
\end{center}
\end{minipage}
\hspace{0.2cm}
\begin{minipage}{3.5cm}
\includegraphics[bb=72 436 540 720,width=3.5cm]{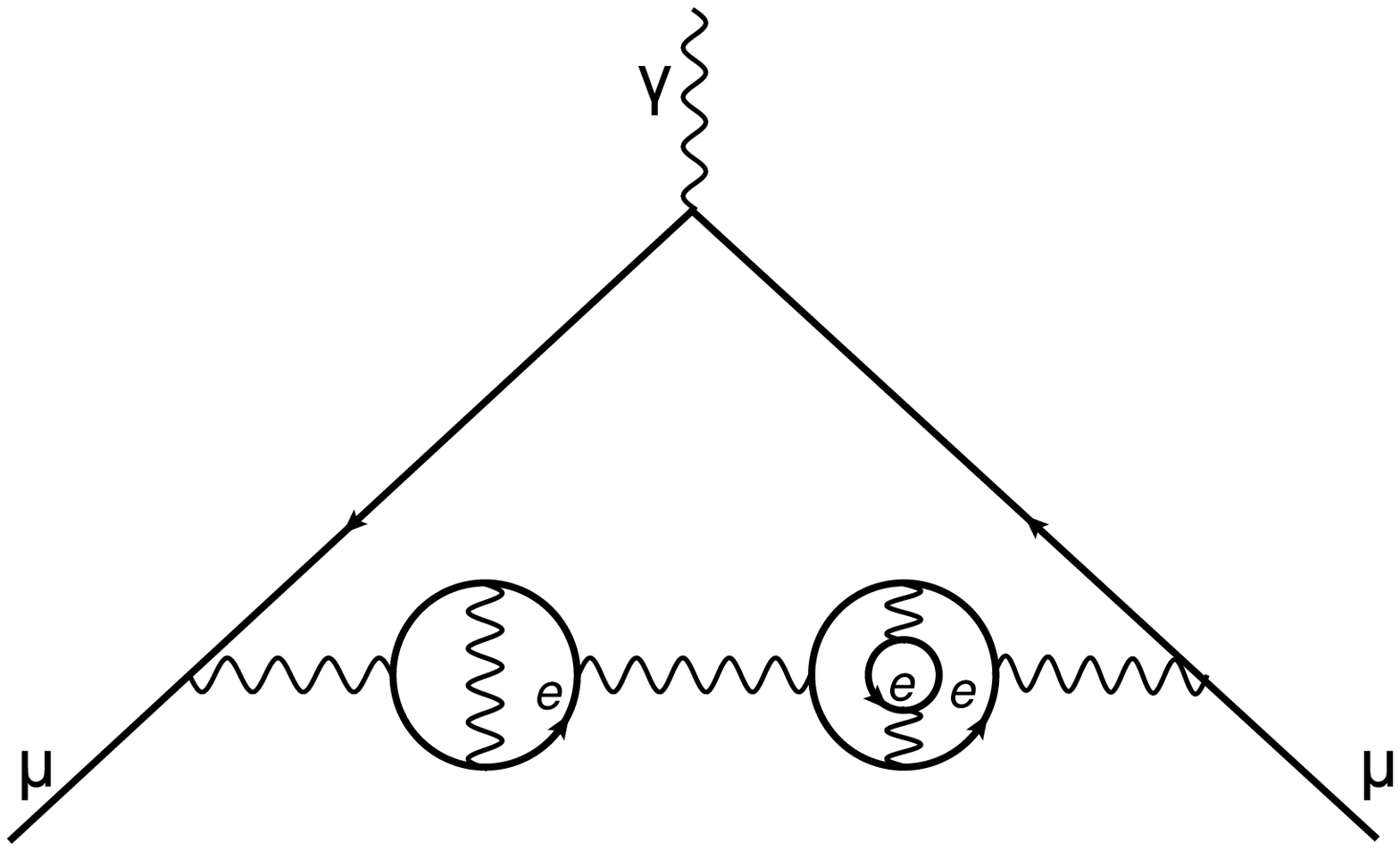}
\begin{center}
\vspace{-0.7cm}
{\tiny$I_6(f)$}
\end{center}
\end{minipage}
\hspace{0.2cm}
\begin{minipage}{3.5cm}
\includegraphics[bb=72 436 540 720,width=3.5cm]{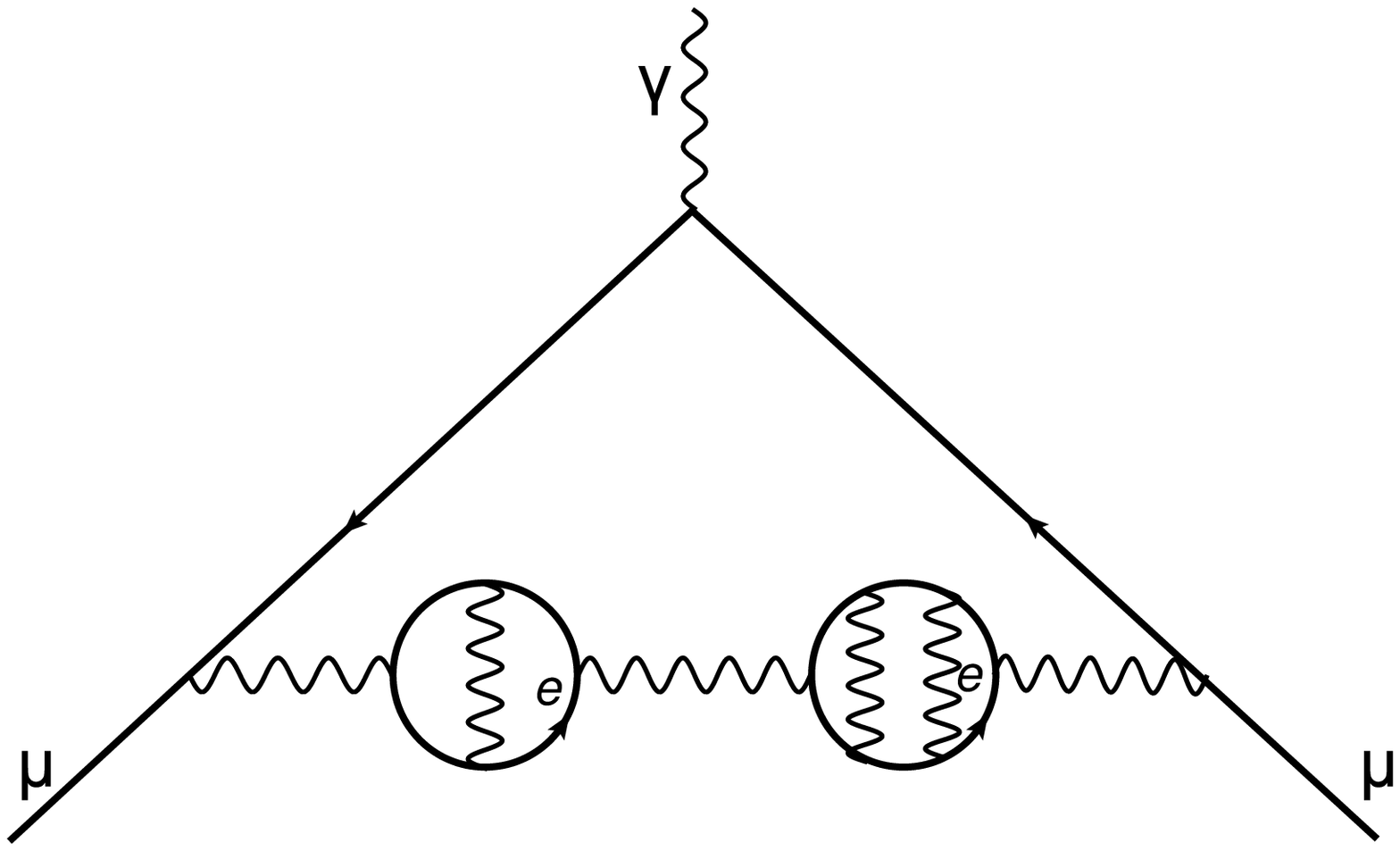}
\begin{center}
\vspace{-0.7cm}
{\tiny$I_6(g)$}
\end{center}
\end{minipage}
\hspace{0.2cm}
\begin{minipage}{3.5cm}
\includegraphics[bb=72 436 540 720,width=3.5cm]{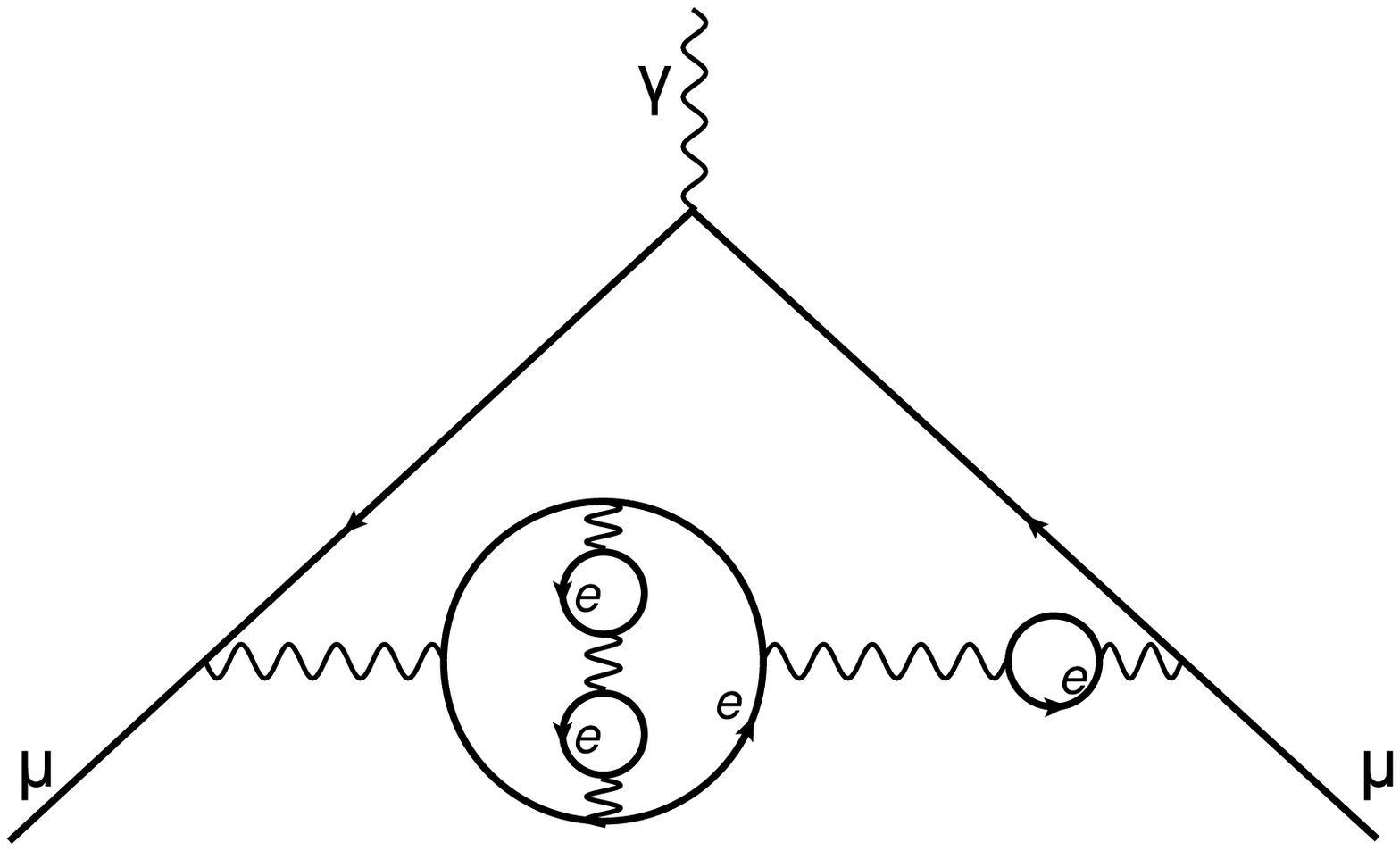}
\begin{center}
\vspace{-0.7cm}
{\tiny$I_6(h)$}
\end{center}
\end{minipage}\\[0.3cm]
\begin{minipage}{3.5cm}
\includegraphics[bb=72 436 540 720,width=3.5cm]{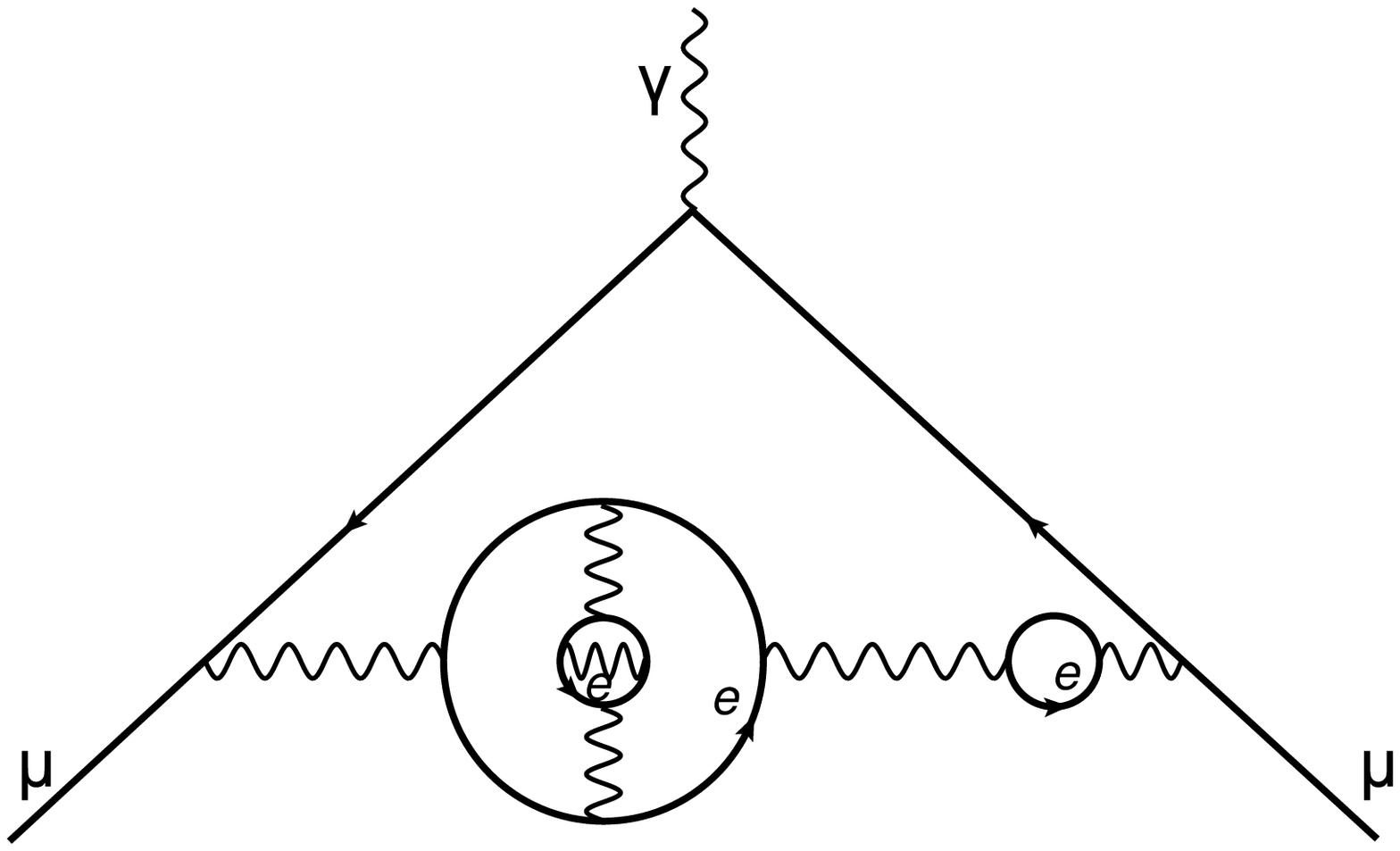}
\begin{center}
\vspace{-0.7cm}
{\tiny$I_6(i)$}
\end{center}
\end{minipage}
\hspace{0.2cm}
\begin{minipage}{3.5cm}
\includegraphics[bb=72 436 540 720,width=3.5cm]{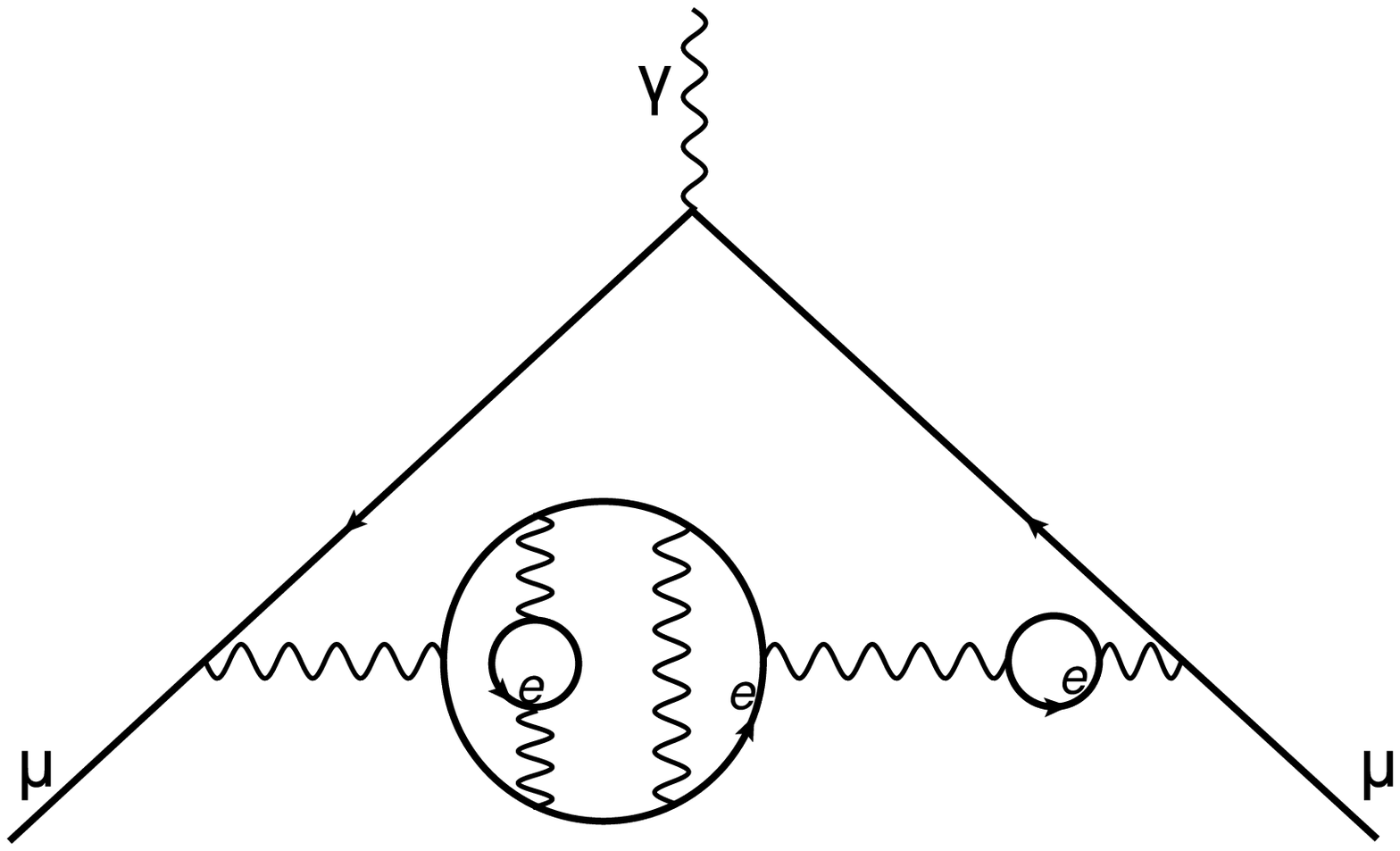}
\begin{center}
\vspace{-0.7cm}
{\tiny$I_6(j)$}
\end{center}
\end{minipage}
\hspace{0.2cm}
\begin{minipage}{3.5cm}
\includegraphics[bb=72 436 540 720,width=3.5cm]{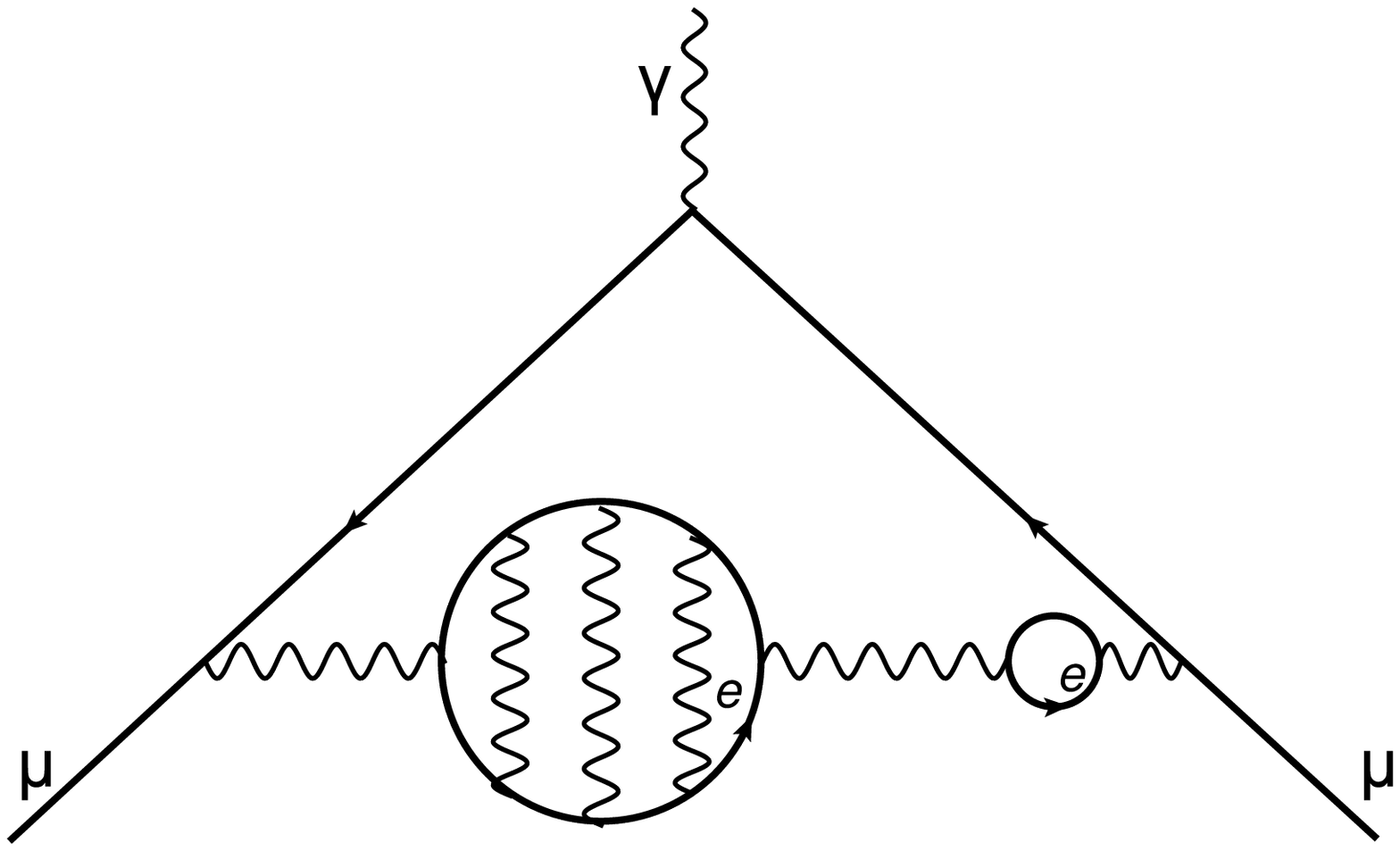}
\begin{center}
\vspace{-0.7cm}
{\tiny$I_6(k)$}
\end{center}
\end{minipage}
\hspace{0.2cm}
\begin{minipage}{3.5cm}
\includegraphics[bb=72 436 540 720,width=3.5cm]{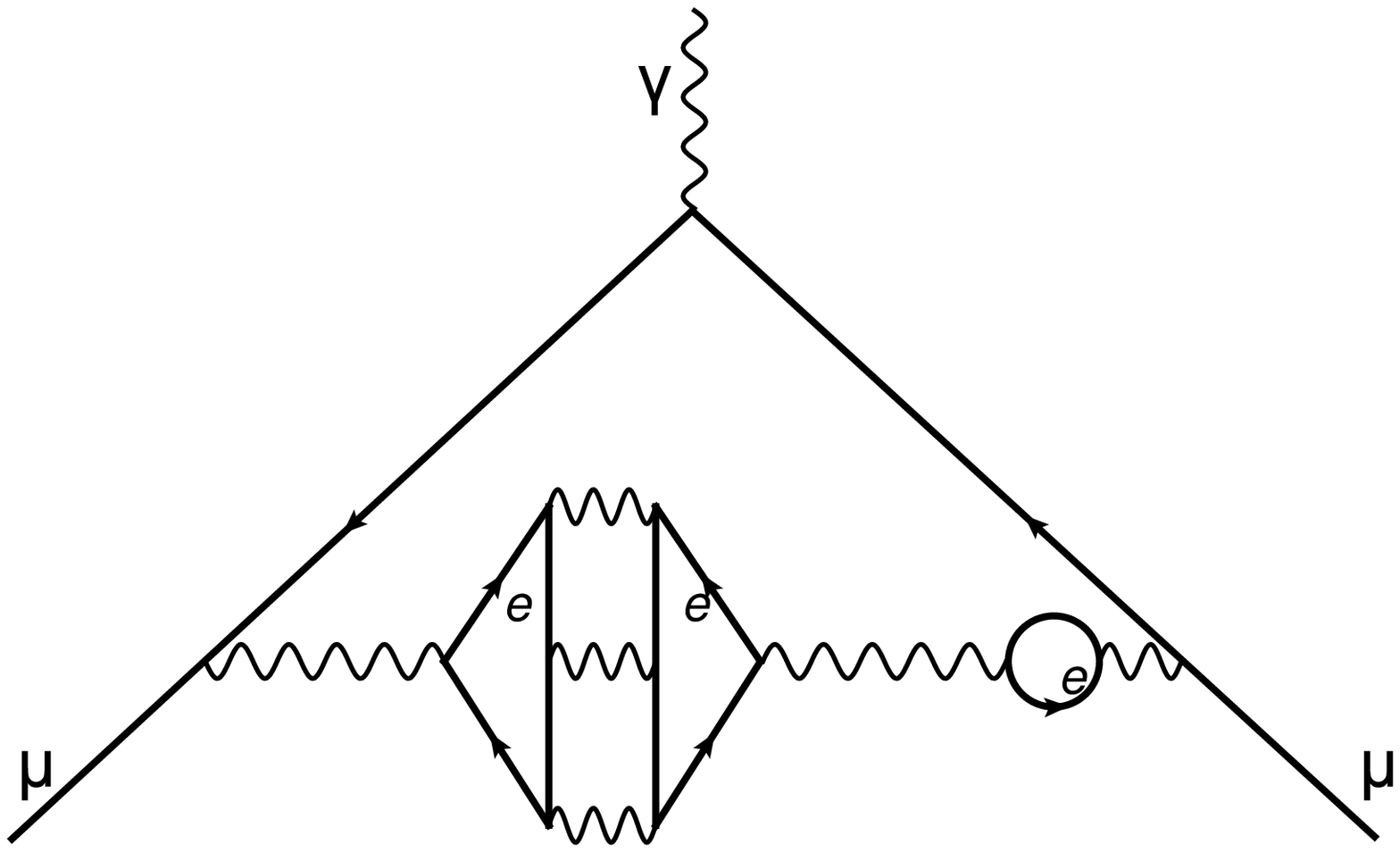}
\begin{center}
\vspace{-0.7cm}
{\tiny$I_6(l)$}
\end{center}
\end{minipage}
\end{center}
\vspace{-0.3cm}
\caption{\label{fig:FactMuonClasses6loop} The different diagram classes
  of the factorizable six-loop contribution. For each diagram class
  $I_6(a)$--$I_6(l)$ one typical representative is shown.}
\end{figure}

The factorizable contribution of Eq.~(\ref{amu6}), $\left.a^{\scriptsize
  asymp,(6)}_\mu\right|_{\mathrm{factr}} =  246.381$, which originates from
the terms in $d_5^{\mathrm{factr}}$ of Eq.~(\ref{eq:d5factr}) can also
be classified into several subsets: 
\begin{eqnarray}
\left.a^{\scriptsize  asymp,(6)}_\mu\right|_{\mathrm{factr}} &=&
 \amu{(6)}{a}+\amu{(6)}{b}+\amu{(6)}{c}+\amu{(6)}{d}+\amu{(6)}{e}+\amu{(6)}{f}
\nonumber\\
&+&\amu{(6)}{g}+\amu{(6)}{h}+\amu{(6)}{i+j}+\amu{(6)}{k}+\amu{(6)}{l}
{},
\label{eq:amu6factdecomp}
\end{eqnarray}
where we have shown in Fig.~\ref{fig:FactMuonClasses6loop} one typical
contributing diagram for each term on the r.h.s. of
Eq.~(\ref{eq:amu6factdecomp}). The letter in the upper case curly
brackets denotes again from which contribution of
Fig.~\ref{fig:FactMuonClasses6loop} the considered term is coming. The
numerical values of the factorizable contributions are shown in
Table~\ref{tab:sixloopfact}.
%
%
\begin{table}[!h]
\begin{center}
\begin{tabular}{|c|c|c|c|c|c|c|}
\hline
{Subset}&$I_{6}(a)$&$I_{6}(b)$&$I_{6}(c)$&$I_{6}(d)$&$I_{6}(e)$&$I_{6}(f)$\\\hline
{Value} & 58.1861  & 101.150 & 35.8953 & 30.6997  & -5.12107 & 9.68427\\\hline\hline
{Subset}&$I_{6}(g)$    &$I_{6}(h)$   &\multicolumn{2}{|c|}{$I_{6}(i)+I_{6}(j)$}&$I_{6}(k)$  &$I_{6}(l)$   \\\hline
{Value} & -1.81937    & 15.6754     &\multicolumn{2}{|c|}{7.73424    }       & 0.68395   & -6.38790\\\hline
\end{tabular}
\end{center}
\caption{\label{tab:sixloopfact} Numerical evaluation for the individual
  factorizable contributions of Eq.~(\ref{eq:amu6factdecomp}). The
  numerical values are correct up to power corrections of the order
  $\mathcal{O}\left(M_e/M_\mu\right)$ and arise from the diagram
  classes shown in Fig.~\ref{fig:FactMuonClasses6loop}.}
\end{table}\\
Finally we convert the fine-structure constant in the perturbative
expansion of the anomalous magnetic moment of the muon 
to the $\MSbar$-scheme by using the conversion
relations of Eqs.(\ref{eq:expCaabar})-(\ref{bsMSfromOS4}).
As in the previous sections we consider here only the higher order
QED corrections due to insertions of the electron vacuum polarization function 
into the leading order muon vertex diagram, as described
in Section~\ref{sec:Introduction} and label the corresponding contribution to the anomalous
magnetic moment with $a^{\mbox{\scriptsize{{eVP}}}}_{\mu}$. Setting the
scale to the mass of the muon we observe that all scale-depended
logarithms vanish and we obtain
\begin{eqnarray}
\label{eq:amums}
a^{\mbox{\scriptsize{eVP}}}_\mu&=& 
   \frac{\alpha_{\mu}}{2\*\pi}
 - \oalpMmu^{\!\!2}\!\*{25\over 36} 
 + \oalpMmu^{\!\!3}\!\*\left(
   {241\over 2592} 
 + {\pi^2\over27} 
 + {\z3\over2}
              \right) 
 + \oalpMmu^{\!\!4}\!\*\left(
   {253225\over 93312} 
\right.\nonumber\\&&\left.
 - {55\over648}\*\pi^2 
 - {257\over 144}\*\z3 
 - {5\over4}\*\z5
              \right) 
 - \oalpMmu^{\!\!5}\!\*\left(
   {63114877\over 6718464}
 + {1361\over 5832}\*\pi^2
\right.\nonumber\\&&\left.
 - {269\over17280}\*\pi^4
 + {1223\over5184}\*\z3
 - {2\over9}\*\pi^2\*\z3 
 - {4\over3}\*\z3^2 
 - {2795\over576}\*\z5 
 - {35\over8}\*\z7
              \right)
 + \mathcal{O}\!\left(\alpha_{\mu}^6\right)\!,\,
\end{eqnarray}
with $\alpha_{\mu}=\overline{\alpha}(\mu=M_{\mu})$. We have checked that
this also holds at six-loop order, where the remaining scale-independent
parts are not shown in Eq.~(\ref{eq:amums}) since they contain still
unknown contributions, for example those which arise from the unknown
constant $C_5$ of Eq.~(\ref{eq:d5decomp}).
%
%
%
%
\section{Summary and conclusion\label{sec:DiscussConclude}}

We have computed analytically the vacuum polarization function at
four-loop order in perturbative QED in the limit of small and large
momentum, respectively.  From the low energy limit we have derived the
conversion factor of the fine structure constant between on-shell
and $\MSbar$ scheme.  These results have been used to derive 
analytical expressions  for the dominant and gauge invariant contributions to
the muon anomaly originating from vacuum polarization function
insertions at five-loop order. Numerical results for these contributions
are already available in the literature and are up to power corrections
in agreement with our analytical ones.

  Using the result of the recently
computed five-loop QED $\beta$-function in the $\MSbar$ scheme of
Ref.~\cite{Baikov:2012zm} we also determine the asymptotic momentum
dependent part of the polarization function at five loops which in turn
allows to calculate vacuum polarization type, asymptotic, leading
contributions to the anomalous magnetic moment of the muon at six loops.
The five-loop momentum dependent part of the polarization function is
also used to determine the five-loop QED $\beta$-function in the
on-shell scheme.\\ 

All the  calculations described here were done during 2007-2008. 
A short version of this  work, dealing only with the N=1 QED was reported at 
the 9th DESY Workshop on Elementary Particle Theory 
in the spring of 2008 \cite{Baikov:2008si}.

Our results will be made available in computer readable form under the
URL: {\tt{http://www-ttp.physik.uni-karlsruhe.de/Progdata/ttp12/ttp12-021                                       }}.

\vspace{2ex}
\noindent
{\bf Acknowledgements}\\
This 
work was supported by
the Deutsche Forschungsgemeinschaft in the
Sonderforschungsbereich/Transregio
SFB/TR-9 ``Computational Particle Physics''
and  by RFBR grants  11-02-01196, 10-02-00525.
This work was also partially supported by U.S. Department
of Energy under contract No.DE-AC02-98CH10886.
The computer  calculations were partially  performed on  the  HP XC4000  supercomputer of  the  federal state Baden-W\"urttemberg at the High Performance Computing Center Stuttgart
(HLRS) under the grant ``ParFORM''.

The figures have been drawn with the help of
Axodraw \cite{Vermaseren:1994je} and JaxoDraw  \cite{Binosi:2003yf}.
\newpage

\begin{appendix}
\section{$\MSbar$--on-shell relation for fermion masses in QED\label{app:MassMSbarOnShell}}
The relation between the $\MSbar$-mass $\ovl{m}(\mu)$ and the on-shell mass
$M$ for $N$ identical fermions in QED is given
by\cite{Chetyrkin:1999qi,Melnikov:2000qh}:%
\begin{eqnarray}
\lefteqn{\ovl{m}( \mu ) = M\,\Bigg(1  } 
\nonumber\\
&{+}&\left(\frac{\ovl{ \alpha}}{\pi} \right)^1
\left[
-1 
-\frac{3}{4} \,\ell_{\mu M}\,
\right]
\nonumber\\
&{+}&\left(\frac{\ovl{ \alpha}}{\pi} \right)^2
\left[
\frac{7}{128} 
+\frac{143}{96}  \, N 
-\frac{5}{16} \,\pi^2
-\frac{1}{6}  \, N \,\pi^2
-\frac{3}{4}  \,\zeta_{3}
+\frac{21}{32} \,\ell_{\mu M}\,
\BreakI
\phantom{+\left(\frac{\ovl{ \alpha}}{\pi} \right)^2}
+\frac{13}{24}  \, N \,\ell_{\mu M}\,
+\frac{9}{32} \,\ell_{\mu M}^2
+\frac{1}{8}  \, N \,\ell_{\mu M}^2
+\frac{1}{2} \,\pi^2 \logtwo \, 
\right]
\nonumber\\
&{+}&\left(\frac{\ovl{ \alpha}}{\pi} \right)^3
\left[
-\frac{2969}{768} 
+\frac{1067}{576}  \, N 
-\frac{9481}{7776}  \, N^2
-\frac{613}{192} \,\pi^2
-\frac{85}{108}  \, N \,\pi^2
+\frac{4}{135}  \, N^2\,\pi^2
\BreakI
\phantom{+\left(\frac{\ovl{ \alpha}}{\pi} \right)^3}
-\frac{1}{48} \,\pi^4
+\frac{91}{2160}  \, N \,\pi^4
-\frac{81}{16}  \,\zeta_{3}
-\frac{53}{24}  \, N  \,\zeta_{3}
+\frac{11}{18}  \, N^2 \,\zeta_{3}
-\frac{1}{16} \,\pi^2 \,\zeta_{3}
\BreakI
\phantom{+\left(\frac{\ovl{ \alpha}}{\pi} \right)^3}
+\frac{5}{8}  \,\zeta_{5}
-\frac{489}{512} \,\ell_{\mu M}\,
-\frac{151}{384}  \, N \,\ell_{\mu M}\,
-\frac{197}{216}  \, N^2\,\ell_{\mu M}\,
+\frac{15}{64} \,\pi^2\,\ell_{\mu M}\,
+\frac{1}{3}  \, N \,\pi^2\,\ell_{\mu M}\,
\BreakI
\phantom{+\left(\frac{\ovl{ \alpha}}{\pi} \right)^3}
+\frac{1}{9}  \, N^2\,\pi^2\,\ell_{\mu M}\,
+\frac{9}{16}  \,\zeta_{3}\,\ell_{\mu M}\,
-\frac{1}{4}  \, N  \,\zeta_{3}\,\ell_{\mu M}\,
-\frac{27}{128} \,\ell_{\mu M}^2
-\frac{13}{32}  \, N \,\ell_{\mu M}^2
\BreakI
\phantom{+\left(\frac{\ovl{ \alpha}}{\pi} \right)^3}
-\frac{13}{72}  \, N^2\,\ell_{\mu M}^2
-\frac{9}{128} \,\ell_{\mu M}^3
-\frac{3}{32}  \, N \,\ell_{\mu M}^3
-\frac{1}{36}  \, N^2\,\ell_{\mu M}^3
+\frac{29}{4} \,\pi^2 \logtwo \, 
\BreakI
\phantom{+\left(\frac{\ovl{ \alpha}}{\pi} \right)^3}
+\frac{8}{9}  \, N \,\pi^2 \logtwo \, 
-\frac{3}{8} \,\pi^2\,\ell_{\mu M}\, \logtwo \, 
-\frac{1}{3}  \, N \,\pi^2\,\ell_{\mu M}\, \logtwo \, 
+\frac{1}{2} \,\pi^2 \Log{2}{2}\,
\BreakI
\phantom{+\left(\frac{\ovl{ \alpha}}{\pi} \right)^3}
-\frac{1}{9}  \, N \,\pi^2 \Log{2}{2}\,
-\frac{1}{2}  \Log{2}{4}\,
+\frac{1}{9}  \, N  \Log{2}{4}\,
-12  \,a_4
+\frac{8}{3}  \, N  \,a_4
\right]
\Bigg)
{},
\label{CmM}
\end{eqnarray}

with $\ell_{{\mu}M}=\ln(\mu^2/M^2)$.

\providecommand{\href}[2]{#2}\begingroup\raggedright\endgroup

\end{appendix}
\end{document}